\documentclass[12pt]{article}
\usepackage{amsfonts}
\usepackage{amsmath}
\usepackage{amsthm} 
\usepackage[usenames]{color}

\usepackage{hyperref}
\usepackage[pdftex]{graphicx}

\newcommand{\NoBlackBoxes}{\global\overfullrule0pt}
\NoBlackBoxes

\newcommand{\E}{{\mathbb E}}
\newcommand{\Var}{{\mathbb Var}}

\newcommand{\Q}{{\mathbb Q}}

\newcommand{\R}{{\mathbb R}}
\newcommand{\N}{{\mathbb N}}

\newcommand{\Fcal}{{\mathcal F}}

\newcommand{\Kcal}{{\mathcal K}}

\newcommand{\Xcal}{{\mathcal X}}
\newcommand{\Ycal}{{\mathcal Y}}

\renewcommand{\omega}{\varpi}

\newcommand{\DA}{{A}}

\newcommand{\one}{{\rm \bf 1}}

\newcommand{\InsertFig}[1]{
    \vspace{5mm}
    \begin{center}
    \textsc{[Figure #1 should be inserted here]}
    \end{center}
    \vspace{5mm} }
    
\newcommand{\InsertTable}[1]{
    \vspace{5mm}
    \begin{center}
    \textsc{[Table #1 should be inserted here]}
    \end{center}
    \vspace{5mm} }

\newcommand{\myfootnote}{\footnote}

\newtheorem{proposition}{Proposition}
\newtheorem{lemma}[proposition]{Lemma}

\newtheorem{remark}[proposition]{Remark}
\newtheorem{exampleemph}[proposition]{Example}   
\newtheorem{foo}[proposition]{Remarks}

\begin{document}

\title{Robust Hedging of Withdrawal Guarantees\\ (Extended Version)\myfootnote{
    The author thanks Alex Langnau (Allianz SE)
    for many useful hints and fruitful discussions that originated this paper. Helpful and inspiring comments
    from two unknown referees and from the colleagues Markus Hirschberger and Richard Vierthauer are
    gratefully acknowledged. }}
\author{Andreas Kunz\myfootnote{Munich Re Group. Letters: K\"oniginstrasse
107, 80802 M\"unchen, Germany. Email: akunz@munichre.com.}
}

\date{Version: \today}

\maketitle

\abstract{
Withdrawal guarantees ensure the periodical deduction of a constant
dollar-amount from a fund investment for a fixed number of periods.
If the fund depletes before the last withdrawal, 
the guarantor has to finance the outstanding withdrawals. We derive
a robust hedging strategy which leads to closed form
solutions for the guarantee value.}

\section{Introduction}
\label{sec:Intro}

In the life insurance industry, withdrawal guarantees have emerged
during the last decades as a guarantee feature in variable annuity
products\myfootnote{Unit-linked life insurance products with
embedded guarantees on the performance of the underlying funds.}.
In typical withdrawal guarantee products, so-called guaranteed
minimum withdrawal benefit (GMWB) policies, the initial
annuitization amount is invested into a fund (mostly with a
significant equity component) and then fixed withdrawal payments are
periodically deducted from the fund. These periodical withdrawals
are guaranteed up to a specified maturity; if the fund depletes
beforehand, the guarantor has to finance the outstanding withdrawal
payments.\myfootnote{If the fund is not depleted at maturity, the
investor (or policy holder) keeps the fund value.}
See \cite{BlamontSagooGMWB, MilevskiGMWB} and the references therein
for further details on GMWB products.

The withdrawal process is conceptually different from consumption
models in standard portfolio optimization approaches, see e.g.
\cite{KaratzasShreve_MethodMathFin_OptContr}, where a constant
fraction of the current fund value is permanently consumed. The
latter consumes less when the fund value has decreased and more when
the fund has outperformed. Withdrawal guarantee products, on the
contrary,  always consume the same absolute dollar-amount, which
leads to a fundamentally different kind of path dependency.

This paper analyzes semi-static hedging strategies for withdrawal
guarantees and uses heavily techniques from  \emph{static hedging}
theory, which was originated by Breeden and Litzenberger
\cite{BreedLitz} and has been  further developed by Carr et al.
\cite{CarrStaticHedgingExoticOption}. The hedging strategies consist
in rolling over 
portfolios of short-dated
put options with different strikes but with common maturity equal to
the subsequent withdrawal time. The hedge portfolios are constructed
to finance 
either all outstanding withdrawals
if the fund depletes
or otherwise the roll-over into the new hedge portfolio
of short-dated put options.

If the market dynamics (here the underlying asset price) is driven
by a one-factor Markov process, the semi-static hedging strategy
leads to a perfect replication. This applies
in particular 
for the local volatility model.\footnote{This  
	model  assumes that instantaneous volatility of the asset price
	depends deterministically on the dynamics of the spot.} 
We show how
the weights of the single put options with specific strikes
 in the hedging portfolio can be derived backward recursively.

If an additional stochastic market factor comes into play, such as
stochastic volatility, then the semi-static hedging strategy
becomes sensitive with respect to moves in this additional factor.
Hence the replication might not be perfect.
We analyze 
in a Black-Scholes setting the forward vega and
volga\footnote{Forward
    vega is the first order and forward volga (or
    vol-convexity) is the second order sensitivity with respect to moves
    in the forward volatility.} sensitivities
of the withdrawal guarantee after forward vega hedge\footnote{with
    forward starting variance swaps}
    and  discover a long volga position
if the guarantee is not too far out-of- or in-the-money. Hence, any move
of the forward volatility works in favor of the hedger.
As stochastic volatility models account for systematic hedging gains
due to the volatility of volatility, they are expected to value
at-the-money withdrawal guarantees at a lower price and vice versa
if the guarantee is far out-of- or in-the-money case. This
expectation is verified for the Heston model by comparing it against
the local volatility model.

If the asset price process has independent returns, i.e. is of
exponential L\'evy type\footnote{a rich class of financial models 
	  contains 
	  Merton's jump diffusion, variance gamma, normal inverse Gaussian or CGMY model, see
    \cite{TankovSurveilExpoLevy} for references and further
    details.},
the expressions for the withdrawal guarantee value and the option
weights in the semi-static hedging strategy simplify considerably.
We show that the option weights for the different time steps can be
represented as density of a process that evolves backwards in time.
This process is identified as value process of a multi-contribution
fund to which constant money amounts are periodically contributed
(rather than deducted as in the withdrawal fund case). This fund is
driven by an asset process that is reverse $\Gamma$-adjoint to the
underlying asset process of the withdrawal guarantee.\footnote{The
    transition kernels of the adjoint asset process driving the multi-contribution fund
    coincide with the Gamma derivatives of
    vanilla options on the original asset process driving the withdrawal fund  in time
    reverse order.}
The withdrawal guarantee value can then be represented as a call
option on the adjoint multi-contribution fund.
The duality between withdrawal guarantee and multi-contribution fund
also holds vice versa: a  put option on a multi-contribution fund is
semi-statically hedgeable and can be interpreted 
as a call option on the
adjoint withdrawal fund.

We also show that the static hedging representation can be extended
to withdrawal guarantees with roll-up feature (also known as ratchet feature)
that increases the
guaranteed withdrawal level when the fund has outperformed.

The paper is organized as follows: section 2 gives a more detailed
description of  withdrawal guarantees. In section
\ref{sec:TwoPeriodCase}, the main ideas for deriving the static
hedging strategy are outlined for the simple two-period case and the
effects of stochastic volatility are analyzed. Section
\ref{sec:MultiPeriodSetting} generalizes for the multi-period case,
exhibits the duality between withdrawal guarantees and
multi-contribution funds and extends to the roll-up case.

\section{Formalizing the Withdrawal Guarantee}
\label{sec:DesciptionGuarantee}

We denote by $X_0$ the initial capital (i.e. the initial annuity
reserve) which is invested into an 
asset with price process $(S_t)_{t \geq 0}$.\myfootnote{We assume a
total return performance for $S$. Hence there is no dividend risk.}
Let $0 = T_0 < T_1 < \dots < T_N =T^*$ denote the schedule of
withdrawal times with maturity $T^*$ and equidistant time intervals
$\Delta T := T^*/N$.
At each time $T_t$, the fixed withdrawal amount $w>0$ is deducted
from the fund.  We denote by $X_t$ the fund value at withdrawal time
$T_t$, $t = 1,\dots, N$, after the withdrawal amount has been
deducted. Analogously, we write short-hand $S_t$ for the asset price
at time $T_t$,
and we assume for simplicity $S_0 = 1$ for the initial asset price.

The fund value process obeys the obvious recursion relation
\begin{equation}\label{eqn:FundValueRecurrion}
    X_{t+1} = X_t \cdot S_{t+1}/S_t - w \, , \qquad t = 0, \dots, N-1\, .
\end{equation}
Iterating the recursion yields 
\begin{equation}\label{eqn:FundValueSolution}
X_{t} = S_t  \cdot \left( X_0 - w\cdot \sum_{u =1}^t \frac{1}{S_u} \right)
    = S_t  \cdot \left( X_0 - \omega \cdot \sum_{u \, : \, 0<T_u \leq T_t} \frac{\Delta T}{S_u} \right)
    \qquad (t = 1, \dots, N)  \, ,
\end{equation}
where $\omega$ denotes the annualized withdrawal rate, i.e. $\omega = w / \Delta T$. 
The sum expression
is known as the harmonic average functional of the asset price $S$. 

The withdrawal guarantee works as follows: the withdrawal payments
until maturity $T^*$ are guaranteed to the investor (or policy
holder) even if the fund depletes before $T^*$. In this case, the
guarantor must finance the outstanding withdrawals.

We denote by $\tau$ the depletion time of the fund, which is the
first time when the fund value is not sufficient to finance the
current withdrawal amount $w$, i.e. $\tau := \inf \{t : X_t \leq
0\}$.\footnote{Alternatively, $\tau$ can be expressed as the first
    hitting time of the level $z = X_0 / \omega$ by the positive
    increasing harmonic average functional, i.e.
    $\tau = \inf \{t : 
    \sum_{u=1}^t \frac{\Delta T}{S_u}> z\}$.
    }

\begin{figure}[!h]
\begin{center}
\includegraphics[width=1.0\textwidth,height = !]{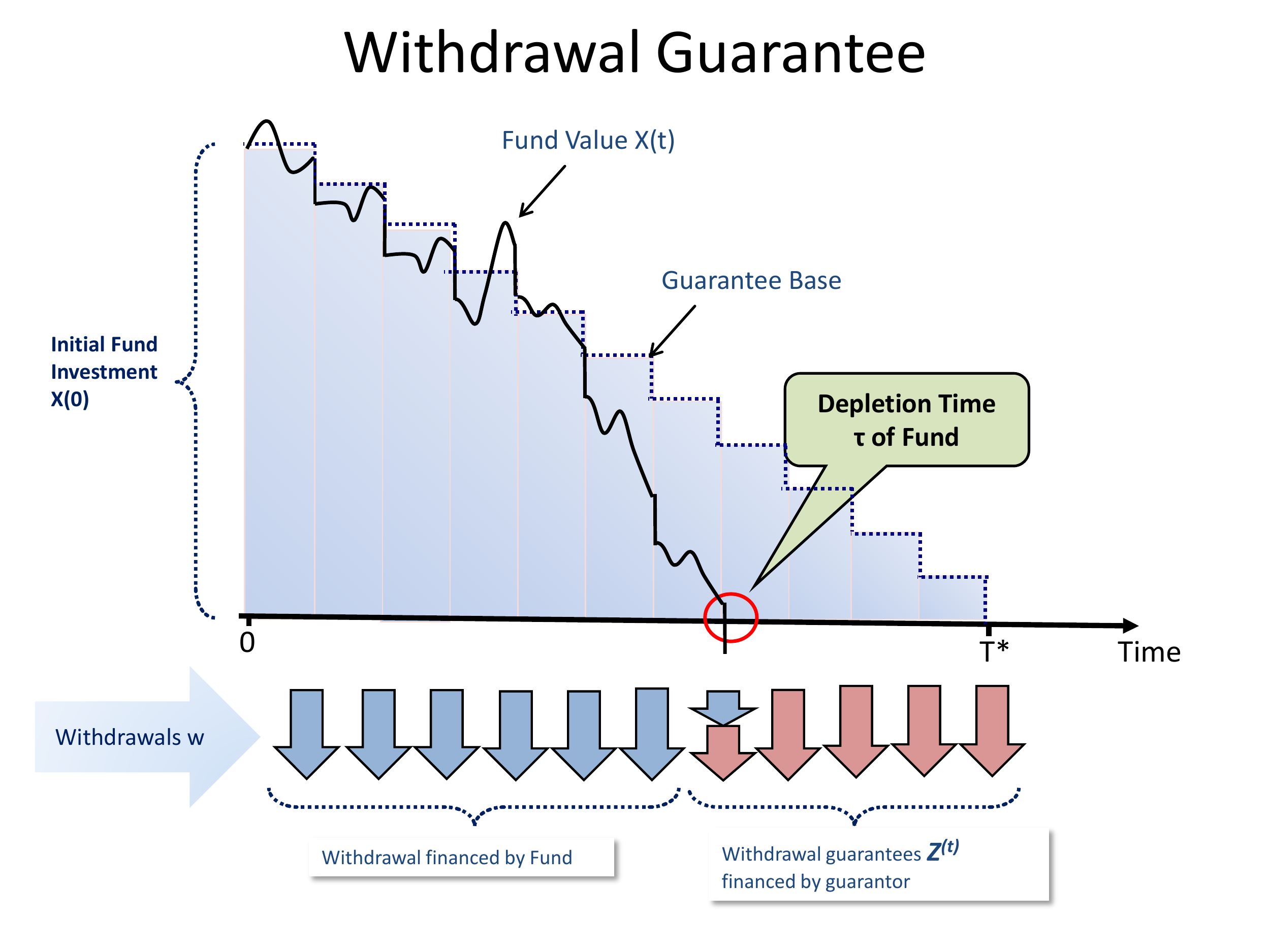}
\caption{\small These graphics illustrate the contingent withdrawal
guarantee payments. The guarantor has to finance the outstanding
withdrawal payments in case the fund is depleted before maturity.
The smaller red arrow 
corresponds to the
expression $X_t^-$ in the representation of the contingent
withdrawal guarantee payment $Z^{(t)}$ due at withdrawal time
$T_t$.} \label{fig:GMWB}
\end{center}
\end{figure}

The total withdrawal guarantee can be decomposed into contingent
guarantee claims $Z^{(t)}$, $t = 1, \dots, N$.
The single claim $Z^{(t)}$ is  due 
at withdrawal time $T_t$ and can be
expressed as follows: if the fund is already depleted before time
$T_t$, the guarantor has to finance the complete withdrawal amount
$w$. If the ruin of the fund happens at time $T_t$, the guarantor
has to  pay only that part of $w$ 
that can no more be financed by the fund, i.e. the negative part
$X_t^- := -\min(0,X_t)$ of the fund value $X_t$ after withdrawal of
the full amount $w$. Hence, $Z^{(t)}$
 can be written 
 as
\begin{eqnarray}
\label{eqn:DefZt}
    Z^{(0)} = 0 \, ,\qquad Z^{(t)} = w \cdot \one_{\tau \leq t-1} + X_t^- \cdot \one_{\tau = t}
    \, , \qquad t = 1, \dots, N\, .
\end{eqnarray}
Here, $\one_{\tau = t}$ denotes the indicator function of the set
$\{\tau = t\}$. The contingent guarantee payments are visualized in
figure \ref{fig:GMWB}.

\InsertFig{\ref{fig:GMWB}}

For simplicity reasons, we assume zero interest
rates.\myfootnote{The extension to non-zero deterministic interest
rates is straightforward.}
 If the market
consistent pricing measure is given by the equivalent martingale
measure $\Q$, then the market value $V_t$ at time $T_t$ of all
guarantee claims payable at time $T_t$ or later reads
\begin{eqnarray}
\label{eqn:defGuaranteeNPVPorcessVt}
    V_t := \E_t^\Q \left[ \sum_{u=t}^N Z^{(u)} \right] \, , \qquad  t = 0, \dots, N\, ,
\end{eqnarray}
where $\E_t^\Q$ denotes conditional expectation with respect to the
filtration $\Fcal_t$ which contains all probability information
observed until time $T_t$.

By inspection of (\ref{eqn:DefZt}) the  value $V_0$ at time $T_0$ of
the total withdrawal  guarantee can also be written as
$$    V_0 = \E^\Q [w\cdot (N-\tau)^+ + X_\tau^- \cdot \one_{\tau \leq
    N}] \, ,$$
and 
reads essentially as a put option on the depletion time $\tau$.

\subsubsection*{Multi-Contribution Fund and Continuous Time Limit}

The withdrawal guarantee is closely related to the dynamics of a
\emph{multi-contribution fund} $Y$ on an underlying asset $\tilde{S}$,
to which a constant money amount
$p>0$ is periodically contributed. 
If the schedule of contributions coincides with that of the
withdrawal case, $Y$ follows a recursion relation analog to
\eqref{eqn:FundValueRecurrion}
\begin{eqnarray}
\label{eqn:DefRecursionMultiContribFund}
     Y_0 := p \, , \qquad Y_{t+1} = Y_t \cdot \tilde{S}_{t+1}/\tilde{S}_t +p  \quad (t = 0, \dots, N-1) \, .
\end{eqnarray}
Iterating this recursion yields
\begin{eqnarray}
\label{eqn:MultiContribFundFormula} Y_{t} = \pi \cdot \tilde{S}_t \cdot
\sum_{u=0}^t \frac{\Delta T}{\tilde{S}_u} \, ,
\end{eqnarray}
where $\pi$ denotes the annualized contribution rate, i.e. $\pi = p/ \Delta T$.

In the continuous time limit, the number of withdrawals $N$ tends to
infinity and $\Delta T \to 0$. It follows immediately from
\eqref{eqn:FundValueSolution} that
the withdrawal fund process $X$ and 
the multi-contribution fund process $Y$
converge 
to the processes
\begin{eqnarray}
\label{eqn:ContinTimeLimit}
    \bar{X}_T =  S_T \cdot \left( X_0 - \omega \int_0^T \frac{du}{S_u} \right) \, , \quad
    \bar{Y}_T =  \pi \cdot \tilde{S}_T \cdot  \int_0^T \frac{du}{\tilde{S}_u}  \qquad (0 \leq T \leq T^*)\, .
\end{eqnarray}
The recursion relations \eqref{eqn:FundValueRecurrion} and \eqref{eqn:DefRecursionMultiContribFund}
translate into the It\^o differential equations
\mbox{$d\bar{X}_t = \frac{\bar{X}_t}{S_t} \, dS_t - \omega \, dt$} and
\mbox{$d\bar{Y}_t = \frac{\bar{Y}_t}{\tilde{S}_t} \, d\tilde{S}_t  + \pi \, dt$}, respectively.

\vspace{5mm}

\emph{Further notations:} 
$X_{t-}:= X_t + w$ denotes the fund value at
time $T_t$ before deduction of the 
withdrawal. Setting 
$\zeta_t := X_t / S_t$, we obtain 
 $X_t = \zeta_{t-1} S_t - w$ and $\zeta_t = \zeta_{t-1} -
w/S_t$ by the recursion relation (\ref{eqn:FundValueRecurrion}).
We denote by $P_{t,t'}(K)$ the (forward) value of a vanilla put
option on $S$  at time $T_{t}$ with maturity $T_{t'}$ and strike
$K$. If 
$T_{t'}=T_{t+1}$, we write $P_{t}(K)$. When we wish
to emphasize the dependence of the option value on the spot value
$S_t$, we write $P_t(K | \,S_t)$. 
We denote by $\Delta_t(K|S_t)$ the first and by $\Gamma_t(K|S_t)$
the second derivative of the put $P_t(K|S_t)$ with respect to the
spot $S_t$, 
the first and second derivative 
in strike dimension is denoted by $\partial_\Kcal P_t(K)$ and
$\partial_\Kcal^2 P_t(K)$.
%

\section{The Two-Period Case}
\label{sec:TwoPeriodCase}

We illustrate the main ideas of hedging withdrawal guarantees in the
two-period case when only two withdrawals are guaranteed at times $T_1$
and $T_2$.

We can rewrite the claim $Z^{(1)}$ due at time $T_1$ as
follows\myfootnote{We assume of course that at time $T_0$ the fund
is not depleted.}
\begin{eqnarray}
\label{eqn:2Period_Z1_Rewrite}
    Z^{(1)} = X_1^- = -\min\left(0, \zeta_0 S_1-w\right) = \zeta_0 \left( w/\zeta_0 -S_1\right)^+ \, .
\end{eqnarray}
Hence $Z^{(1)}$ could be perfectly hedged as of time $T_0$ by
purchasing $\zeta_0$ units of the put $P_0(w/\zeta_0)$.

Hedging the claim $Z^{(2)}$ due at time $T_2$ in addition to
$Z^{(1)}$ is more challenging. We will work backwards in time. At
time $T_1$, the asset price $S_1$ and hence also the fund value
$X_1$ is known. In particular, it is clear whether the fund is
already depleted or not.
Depletion has happened if $X_1 < 0$, or equivalently $S_1 < w /
\zeta_0$. In this case, the guarantor knows that they need to pay $w
- \zeta_0 S_1$ at time $T_1$ and  one period later,  at $T_2$, the
full withdrawals amount $w$, in total the amount $2w - \zeta_0 S_1$.
If the fund is not depleted, the guarantor must buy protection for
the case that the fund at time $T_2$ can not fully finance the
second withdrawal. This protection consists in purchasing $\zeta_1$
units of the put $P_1(w /\zeta_1 | S_1)$ at time $T_1$ which follows
from rewriting the claim $Z^{(2)}$ analogously to
(\ref{eqn:2Period_Z1_Rewrite}).

\begin{figure}[!h]
\begin{center}
\includegraphics[width=0.8\textwidth,height = 0.4\textheight]{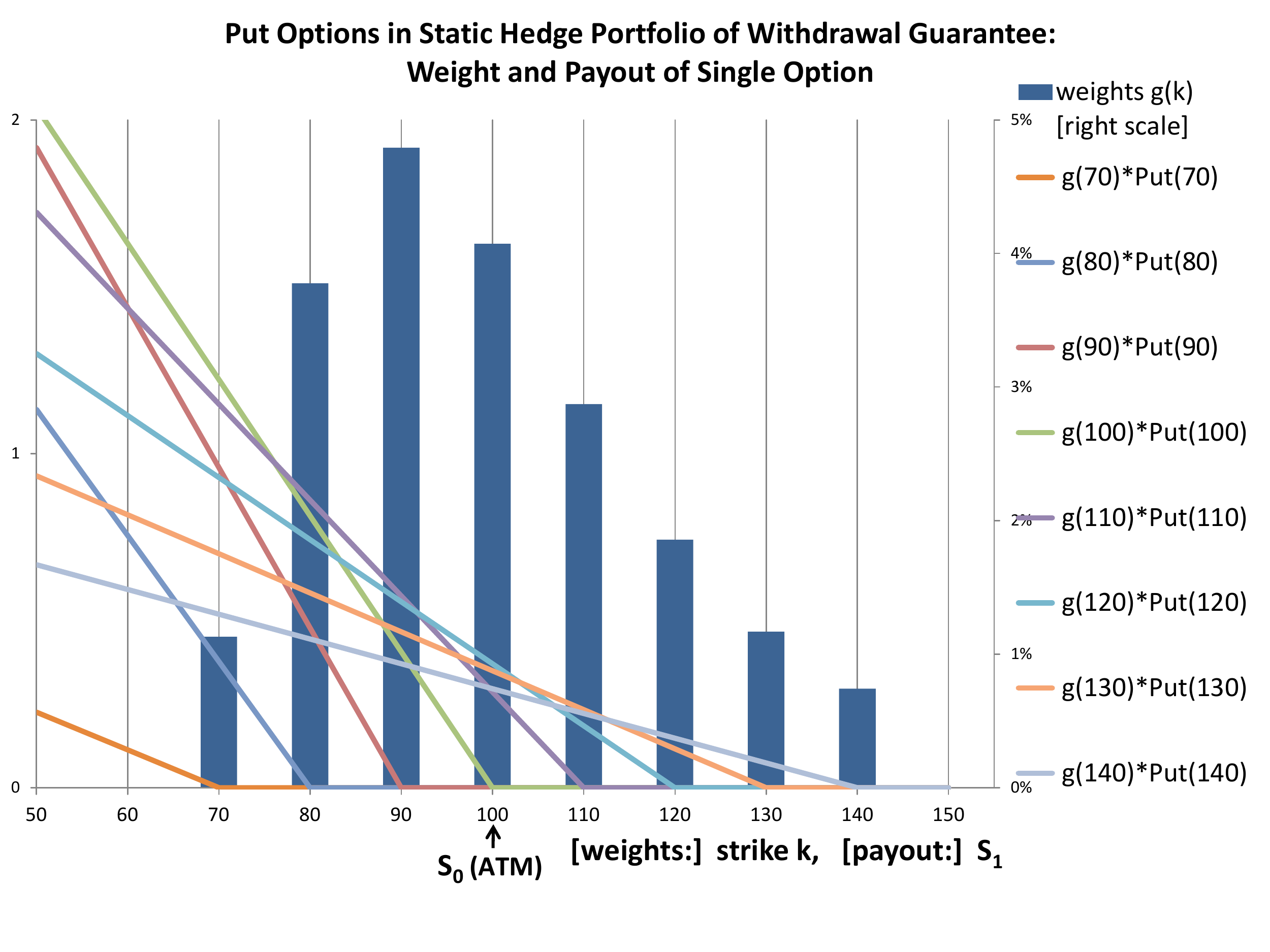}
\includegraphics[width=0.8\textwidth,height = 0.4\textheight]{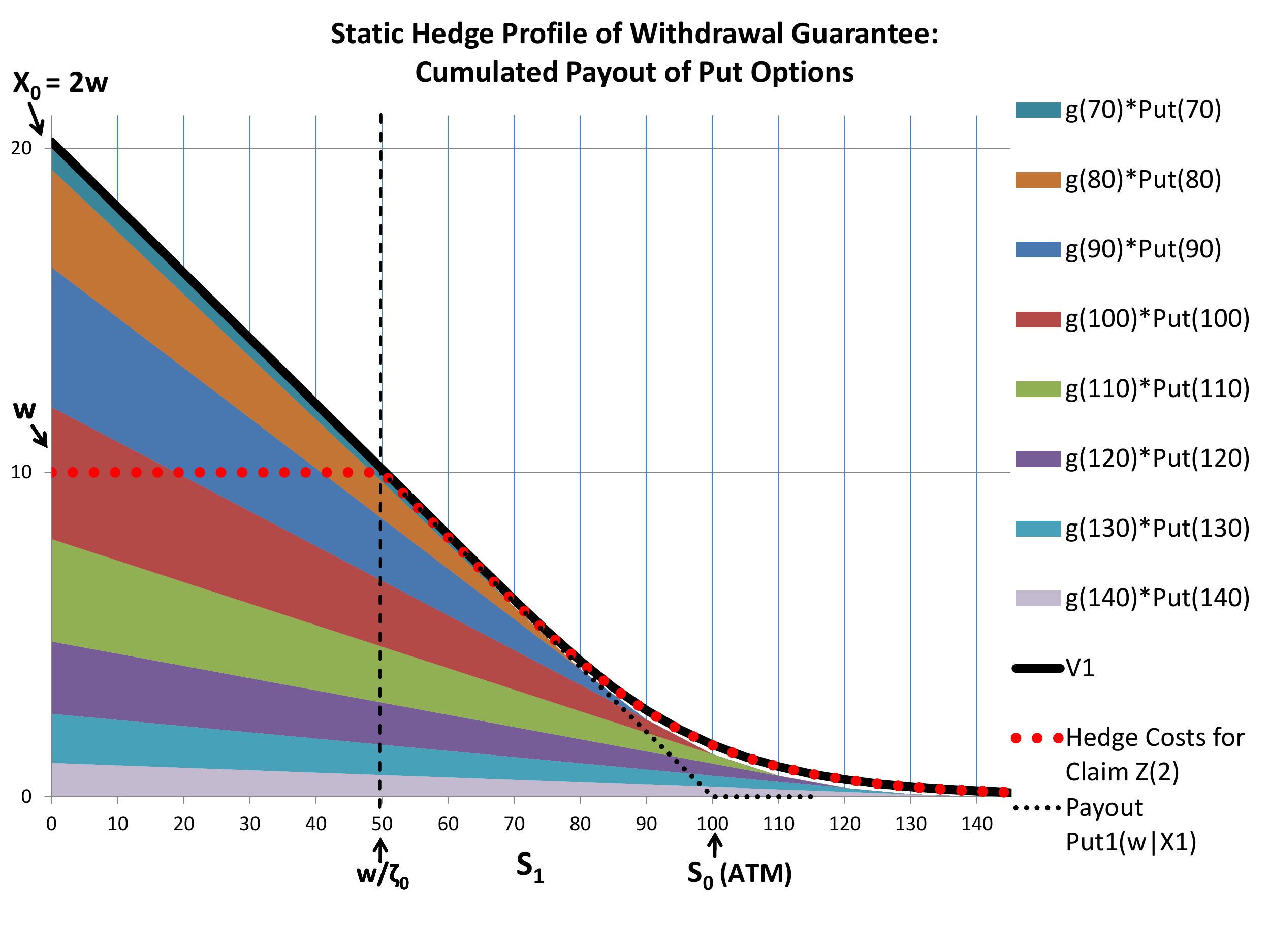}
\caption{ 
Discrete version of the static hedge portfolio for the two-period
withdrawal guarantee in a Black-Scholes setting with volatility set
to $0.25$. Top: weights $g(k)$ of the single put option $P_0(k)$ and
payout profile of the weighted options. Bottom: cumulated payout of
the weighted put options, which matches closely the profile of the
hedging costs $V_1 = V_1(S_1)$ at time $T_1$ (black solid line). The
red dotted line shows only that part of $V_1$ which corresponds to
the hedging costs for the claim $Z^{(2)}$. }
\label{fig:2Period_S1_mapsto_V1}
\end{center}
\end{figure}

Hence the value $V_1$ at time $T_1$ of the outstanding guarantee
claims, i.e. the costs to finance $Z^{(1)}$ and the protection
against 
$Z^{(2)}$ one period later, can be summarized as
\begin{eqnarray}
\label{eqn:2Period_V1}
    V_1 = \left\{
        \begin{array}{lll}
        2w - \zeta_0 S_1&& \mbox{if  }S_1 \leq w /\zeta_0 \, ,\\
        \zeta_1 P_1\left(w/\zeta_1 | S_1 \right)
          = (\zeta_0-\frac{w}{S_1}) \cdot P_1\left(\frac{w}{\zeta_0-\frac{w}{S_1}} \, \big| \, S_1\right)
          && \mbox{if  }S_1 > w /\zeta_0 \, .\\
        \end{array}
        \right.
\end{eqnarray}
The function $V_1$ is visualized by the {black solid line in
the bottom graph} of figure \ref{fig:2Period_S1_mapsto_V1}. The
{red dotted line} shows only that part of $V_1$ which
corresponds to the hedging costs at time $T_1$ for the claim
$Z^{(2)}$.
Note that the function $V_1$ is continuous at $S_1 = w
/\zeta_0$.

\InsertFig{\ref{fig:2Period_S1_mapsto_V1}}

\subsubsection*{Static Hedging} 

The withdrawal guarantee would be perfectly hedged if
the costs for the put option in $V_1$
could be locked in at time $T_0$ already. However, it is generally
not possible 
to perfectly hedge the forward put
$P_1(w/\zeta_1 \, | S_1)$ as of today using vanilla hedging
instruments, since the put 
depends on the future implied volatility at time $T_1$ with
moneyness $(w/\zeta_1)/S_1 = w/X_1$, which is not fixed before time $T_1$.

If the value of the forward put at time $T_1$ depends only on the asset price $S_1$,
then classical static hedging theory can be applied.

Let us recall the famous Breeden-Litzenberger formula: the risk
neutral density of the asset price $S_1$ can be inferred from market
quotes by differentiating vanilla put option prices twice with
respect to the strike, i.e. $\Q(S_1 \in dk) = \partial_\Kcal^2
P_0(k) dk$.\myfootnote{The same works with vanilla call options as
well, of course. Here, $\Q(S_1 \in dk)$ is short-hand writing for
\mbox{$\Q(k \leq S_1 < k +dk)$}.} Hence, any European option with
payout $\phi(S_1)$ at time
$T_1$ can be statically hedged by a portfolio of vanilla put 
options with different strikes. If the function $\phi$ is twice differentiable,
the static hedge portfolio\footnote{The
    real-life hedge portfolio would of course approximate the
    integral by a sum of put options, e.g. $ \sum_{i}
    \delta\phi''(i\delta) \, P_0(i\delta)$, provided the boundary terms
    vanish.}
is obtained by applying twice partial integration on
the option value:
\begin{eqnarray}
    \E^\Q[\phi(S_1)] &=& \int_0^\infty \phi(k) \, \Q(S_1 \in dk) = \int_0^\infty \phi(k) \, \partial_\Kcal^2 P_0(k) dk \nonumber\\
    &=&  \left[ \phi(k) \partial_\Kcal P_0(k) \right]_{k=0}^\infty
    - \left[ \phi'(k)  P_0(k) \right]_{k=0}^\infty + \int_0^\infty \phi''(k) \, P_0(k)
    dk \, .
\label{eqn:2Period_StaticHedging}
\end{eqnarray}

To ensure the prerequisites for static hedging, the following
assumptions are imposed on the asset price process, which are formulated in the
more general multi-period setting:
\begin{description}
  \item[(A1)] 
  $(S_t)_{t\geq 0}$ is a one-factor Markov process and every $S_t$ 
  is continuously distributed\footnote{i.e.
        has no singular values (except eventually from zero).}.
\end{description}
Clearly, all one-dimensional diffusion models for the asset price dynamics 
(Black-Scholes, Dupire's local volatility model, etc.) satisfy assumption (A1).
The same is true for one-dimensional exponential L\'{e}vy models (such as
Merton's jump diffusion, variance gamma, normal inverse Gaussian or CGMY model),
see \cite{TankovSurveilExpoLevy} for references and further details.
Note that assumption (A1) rules out stochastic volatility models.

The Markov property 
implies that forward put options
at time $T_t$ written on $(S_t)_{t\geq 0}$
depend only on the state of the market at time $T_t$
and hence only on the spot $S_t$ which is the unique market factor.
Further, the regularity assumption implies that the put values $P_{t}(K|S_{t})$ are
twice differentiable with respect to strike $K$ (and spot $S_{t}$) for every $t = 0, \dots, N-1$.

Assumption (A1)
ensures that ${V}_1$ depends only on the asset
price $S_1$ at time $T_1$. Hence 
the static hedging formula \eqref{eqn:2Period_StaticHedging} can be applied
to  the function ${V}_1 = {V}_1(S_1)$.
We obtain the hedging strategy
that locks in the guarantee costs $V_1$ already as of time $T_0$ and
hence perfectly replicates the withdrawal guarantee.

The following technical smoothness and boundary conditions on the
function $V_1$ and its first derivative $V_1' := \partial_{S_1} V_1$
guarantee that the boundary terms in the static hedging formula
(\ref{eqn:2Period_StaticHedging}) vanish (the proof of these
conditions has been transferred to the appendix):\footnote{We
    write   $V(k) = o\left(1/k\right)$ as $k\to \infty$ if $V(k)\cdot k \to 0$.
    }
\begin{eqnarray}
\label{eqn:2Period_TechCondit_BdTermVanish}
\mbox{${V}_1$ and ${V}_1'$ are continuous at $w /\zeta_0$ and take the following boundary values:}
\end{eqnarray}
\vspace{-10mm}
\begin{table*}[h!]
\centering
{\small
\begin{tabular}{l|c c}
& ${V}_1(k)$ & ${V}_1'(k)$ \\
\hline
$k = 0$ & $2w$ & $-1$ \\
$k \to \infty$ & $0$ & $o(1/k)$ \\
\end{tabular}}
\end{table*}

This condition 
implies that all boundary terms in the static hedging formula
(\ref{eqn:2Period_StaticHedging}) vanish when we recall the boundary
behavior of the put option 
$P_0(0) = 0$ and $P_0(k) \sim k$, $\partial_\Kcal P_0(k) \to 1$ as
$k\to \infty$.\footnote{We
    write $P(k) \sim k$ as $k\to \infty$ if
    $P(k)/ k \to 1$. }

Keeping in mind that the function ${V}_1$ is linear on the interval $[0,w/\zeta_0]$
with zero second order derivative, the static hedging formula (\ref{eqn:2Period_StaticHedging}) leads to
the following \emph{result}:

Under assumption (A1), the two-period withdrawal guarantee can be
perfectly hedged as of time $T_0$ by the portfolio
\begin{eqnarray}
\label{eqn:2Period_Theorem_HedgePF_V0Repres}
V_0 = \int_{w/\zeta_0}^\infty  \gamma(k)   \, P_0(k) dk
\end{eqnarray}
of put options $P_0(k)$ with common maturity $T_1$ but different strikes $k>0$ that
are weighted according to the function
\begin{eqnarray}
\label{eqn:2Period_Theorem_HedgePF}
\gamma(k) := V_1''(k)=
\left[D_{S_1}^2 \left\{ \left(\mbox{$\zeta_0-\frac{w}{S_1}$}\right)
\cdot P_1\Big(\frac{w}{\zeta_0-\frac{w}{S_1}} \, \big| \, S_1\Big)
\right\}\right]_{S_1 = k} \, , \qquad k > w /\zeta_0 \, .
\end{eqnarray}
The value of the two-period withdrawal guarantee at time $T_0$ is
hence given by $V_0$.

A discrete real-world version of the static hedge portfolio in a Black-Scholes setting is
visualized in figure \ref{fig:2Period_S1_mapsto_V1}. The top graph
displays the weights of the single put option $P_0(k)$ for strikes
according to moneyness levels from $0.7$ to $1.4$, 
together with the payout profile of the
correctly weighted put options. The lower graph shows the cumulated
payout of the  weighted put options and demonstrates that it 
matches closely the hedging costs $V_1$ at time
$T_1$.

\subsubsection*{One-factor Exponential L\'evy Case} 

The static hedging representation
simplifies significantly if we assume that the asset-price process
has independent returns, i.e. $S_{t+1} / S_t$ is independent of
$S_t$ for every $t = 1, \dots, N$. This assumption is equivalent to
the following scaling property\footnote{See
    Lemma
    \ref{lemma:EquivLevyScalinProp} for a proof of this equivalence.
    }
\begin{description}
  \item[(A2)] $P_t(\alpha K | \alpha S_t) = \alpha P_t(K | S_t)$ for
  every $\alpha > 0$ and every $t = 0, \dots, N-1$ and $K>0$.
\end{description}
Of course, one-factor exponential L\'{e}vy models satisfy this assumption.
We do not need the 
assumption of stationary returns; e.g. a Black-Scholes model with deterministic
time-dependent volatility would also be allowed.

Under assumption (A2), the function $V_1$ in \eqref{eqn:2Period_V1}
collapses to $V_1(S_1) =  P_1\left({w}\, | \, \zeta_0 S_1-{w}
\right)$ for  \mbox{$S_1 > w/\zeta_0$}.
Hence the weight function $\gamma$ in \eqref{eqn:2Period_Theorem_HedgePF} can be
rewritten:
$$\gamma(k) = V_1''(k) = \zeta_0^2  \cdot \Gamma_1\left({w}\, | \, \zeta_0 k-{w} \right)
       = \zeta_0^2   \cdot  g\left( \zeta_0 k\right) \, , \qquad k > w /\zeta_0 \, ,$$
with function $g$ defined as
\begin{eqnarray}
\label{eqn:2Period_BS_g}
g\left( k'\right) := \Gamma_1\left({w}\, | \, k'-{w} \right)  \, ,\qquad k' > w\, .
\end{eqnarray}

Under the additional assumption (A2), the hedge portfolio for the
two-period withdrawal guarantee in
\eqref{eqn:2Period_Theorem_HedgePF_V0Repres} hence simplifies as
follows (when applying a change of integration variable $k' =
\zeta_0 k$):
\begin{eqnarray}
\label{eqn:2Period_BS_V0}
V_0 &=&  \int_{w /\zeta_0}^\infty
\zeta_0^2 \cdot g\left( \zeta_0 k\right) \cdot P_0(k \, | \, S_0) \, dk
  =\int_{w}^\infty g(k') \cdot P_0(k' \, | \, X_0)  \, dk' \,.
\end{eqnarray}

\subsubsection*{Effects of Stochastic Volatility}

It is evident from the perfect static hedging replication in \eqref{eqn:2Period_BS_V0}
that the sensitivity
of the withdrawal guarantee 
with respect to
instantaneous shocks of the asset price (delta and gamma) stems
exclusively from the portfolio of short-term put options $P_0(k)$,
whereas the weight function  $g$ does not contribute.
The story is different for the sensitivity with respect to
volatility shocks (vega). While the short-term vega stemming from
shocks of the implied volatility with maturity $T_1$ is statically
hedged by the portfolio of short-term puts $P_0(k)$, the forward put
$P_1(w /\zeta_1 \, |\,  S_1)$ and hence the weight function $g$
depend on the $T_1$-in-$T_2$-forward
volatility\myfootnote{$\Sigma_1$ is formally defined by the relation
$(T_2-T_1)\cdot\Sigma_1^2 = \Var[\ln S_2 ] - \Var[\ln S_1]$.}
$\Sigma_1$, which leads to a forward-vega exposure.

If the volatility is allowed to be stochastic, i.e. if the
assumption (A1) is dropped, the forward-vega exposure from the
weight function $g$ must also be hedged. It is intuitive to hedge
this exposure with a $T_1$-in-$T_2$-forward variance swap which is
only sensitive
with respect to the forward volatility $\Sigma_1$.

The variance swap hedge is constructed to offset the first-order
sensitivity of the withdrawal guarantee with respect to changes in
forward volatility, but the second order forward volatility
sensitivity (known as vega-gamma or volga) of the net position after
hedge does not necessarily vanish. When forward volatility moves, a
long volga net position works in favor of the hedger and a short
volga position would produce systematic losses.\myfootnote{Long or
short volga means a position with positive or negative volga,
respectively.}
Since stochastic volatility models typically account for these systematic
gains or losses, they are expected to produce higher model values for guarantees with
dominant short volga net position and
vice versa for a long volga position.

In the remaining part of this section, we analyze in a Black-Scholes
setting the volga position of the withdrawal guarantee after the
forward variance swap hedge and observe 
long and short volga positions depending on the
moneyness\myfootnote{The moneyness of the withdrawal guarantee is
defined by the ratio between the sum of all guaranteed withdrawal
amounts and initial value $X_0$.}. 
We will then compare these findings with the \mbox{Heston} model. In
particular, we analyze for different moneyness levels of the
withdrawal guarantee the model value differences of the Heston and
the corresponding local volatility model.

In the Black-Scholes world,
the forward-vega  $\partial_{\Sigma_1} V_0$ and
forward-volga exposure $\partial_{\Sigma_1}^2 V_0$ of the withdrawal
guarantee can be expressed in terms of Black-Scholes formulae\myfootnote{The gamma
term in the
    weight function (\ref{eqn:2Period_BS_g}) reads
    $\Gamma_1(\kappa|x) = n(d_1)/(x\Sigma_1)$ with $\kappa = w$ and $x=k-w$, where $n$ denotes the density
    of the standard normal distribution and $d_{1/2} = \ln(x/\kappa)/\Sigma_1
    \pm \Sigma_1/2$.
    The forward-vega $\partial_{\Sigma_1} V_0$ or forward-volga
    can be obtained by replacing $\Gamma_1$ in
    (\ref{eqn:2Period_BS_V0})
    by $\partial_{\Sigma_1} \Gamma_1 =
    (d_1 d_2 -1) \Gamma_1/\Sigma_1$ or by $\partial_{\Sigma_1}^2
    \Gamma_1 =  [2+(d_1 d_2)^2 - 3d_1 d_2 - d_1^2 -
    d_2^2]\Gamma_1/\Sigma_1^2$, respectively.}
by means of equations \eqref{eqn:2Period_BS_g} and \eqref{eqn:2Period_BS_V0}.
The forward-vega exposure of the hedge with $T_1$-in-$T_2$-forward
variance swaps must match the exposure of the withdrawals guarantee,
i.e. must be equal to $\partial_{\Sigma_1} V_0$. Since the payout of
a variance swap is quadratic in volatility, its vega is linear in volatility
and volga is constant and equals 
$\mbox{vega} \, / \, \mbox{volatility}$.
Hence the forward-volga of the forward
variance swap hedge for the withdrawal guarantee is given by
$\partial_{\Sigma_1} V_0 /\Sigma_1$ and the forward-volga of the net
position after hedge reads $\partial_{\Sigma_1} V_0 /\Sigma_1 -
\partial_{\Sigma_1}^2 V_0$.

\begin{figure}[!h]
\begin{center}
\includegraphics[width=0.8\textwidth,height = 0.4\textheight]{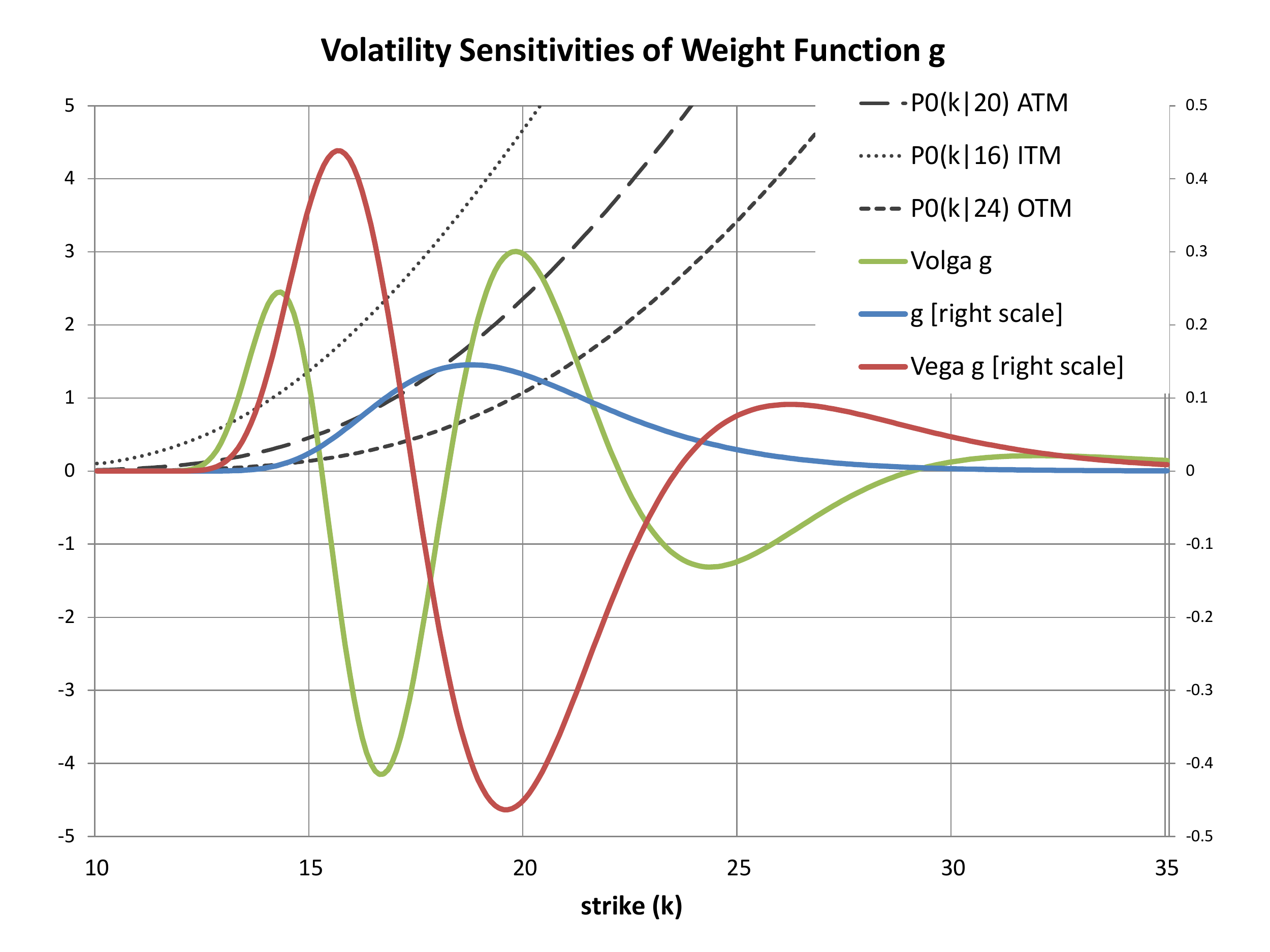}
\includegraphics[width=0.8\textwidth,height = 0.4\textheight]{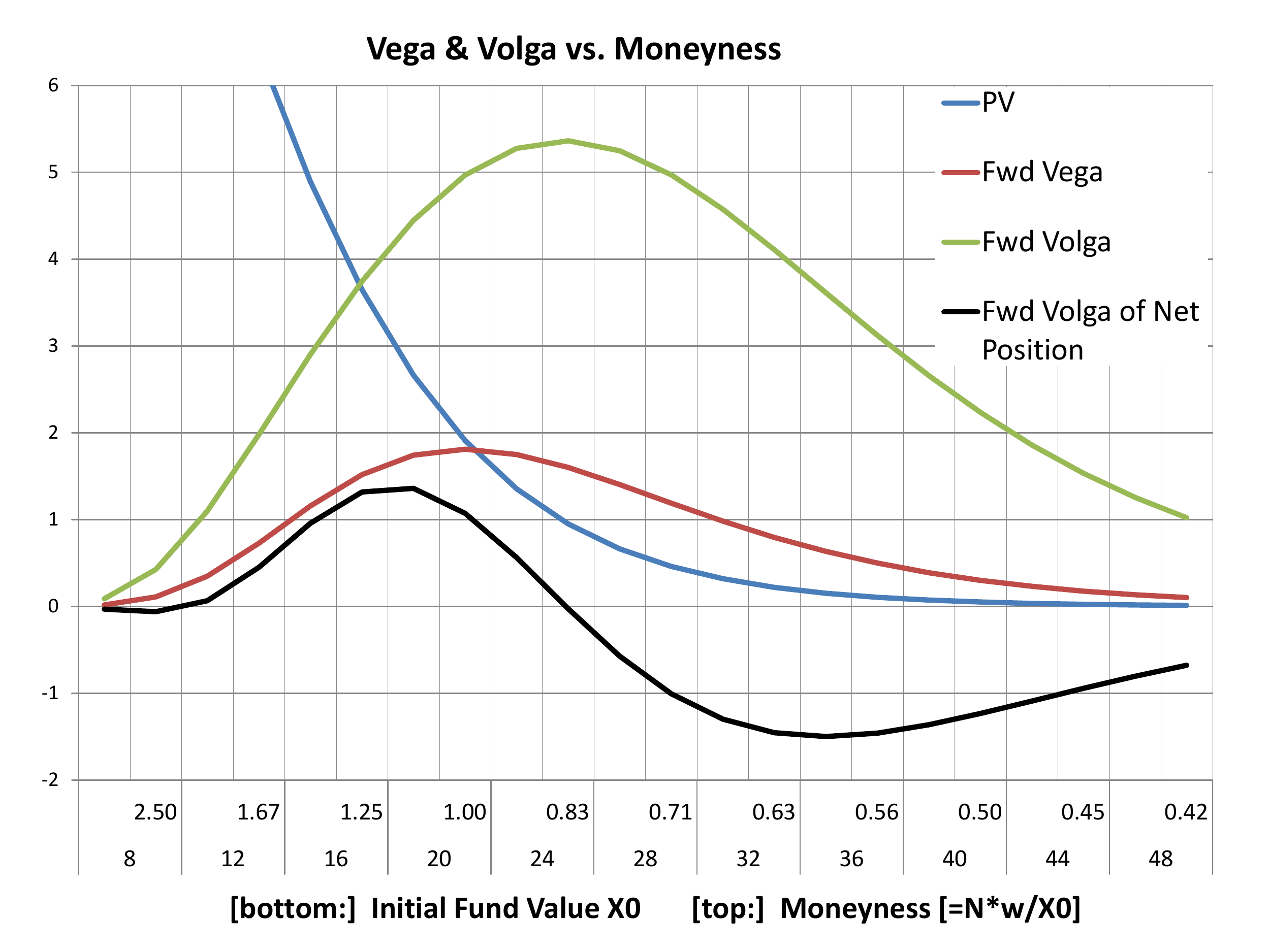}
\caption{The top graph shows the weight function $g$ in the representation of the two-period withdrawal guarantee
and its sensitivities (vega and volga) with respect to the $T_1$-$T_2$-forward volatility. Additionally, put values as a function of strike for different initial fund values are shown.
Integrating the weight function or its volatility sensitivities against the put value functions
gives the corresponding withdrawal guarantee values, which are shown in the bottom graph
for different moneyness levels. In addition, the volga of the net position after hedge with forward
starting variance swap is shown.
Setting: Black-Scholes model with volatility $=0.3$, zero interest rates and dividends;
withdrawal amount $w=10$.}
\label{fig:2Period_VegaVolga_BS}
\end{center}
\end{figure}

\InsertFig{\ref{fig:2Period_VegaVolga_BS}}

In the top graph in figure \ref{fig:2Period_VegaVolga_BS}, the
weight function $g$ defined in \eqref{eqn:2Period_BS_g}
is displayed together with its first and second derivative
$\partial_{\Sigma_1} g$ and $\partial_{\Sigma_1}^2 g$ with respect
to the forward volatility $\Sigma_1$. As outlined in
\eqref{eqn:2Period_BS_V0}, 
we obtain the present value as well as the
forward-vega and forward-volga exposure of the withdrawal guarantee
by integrating these functions against the initial put
prices $k \mapsto P_0(k | X_0)$. 
The bottom graph in figure \ref{fig:2Period_VegaVolga_BS} visualizes
these values for different initial fund values or moneyness levels,
respectively,
together with the volga of the net position after hedge.
A long volga net position is observed for moneyness levels in a wide range around
the at-the-money case. 
In the extreme out-of-the-money case, when the sum of the guaranteed
amounts is less than ca. 70\% of the initial capital,
a short volga position shows up.

\begin{figure}[!h]
\begin{center}
\includegraphics[width=1.0\textwidth,height = 0.4\textheight]{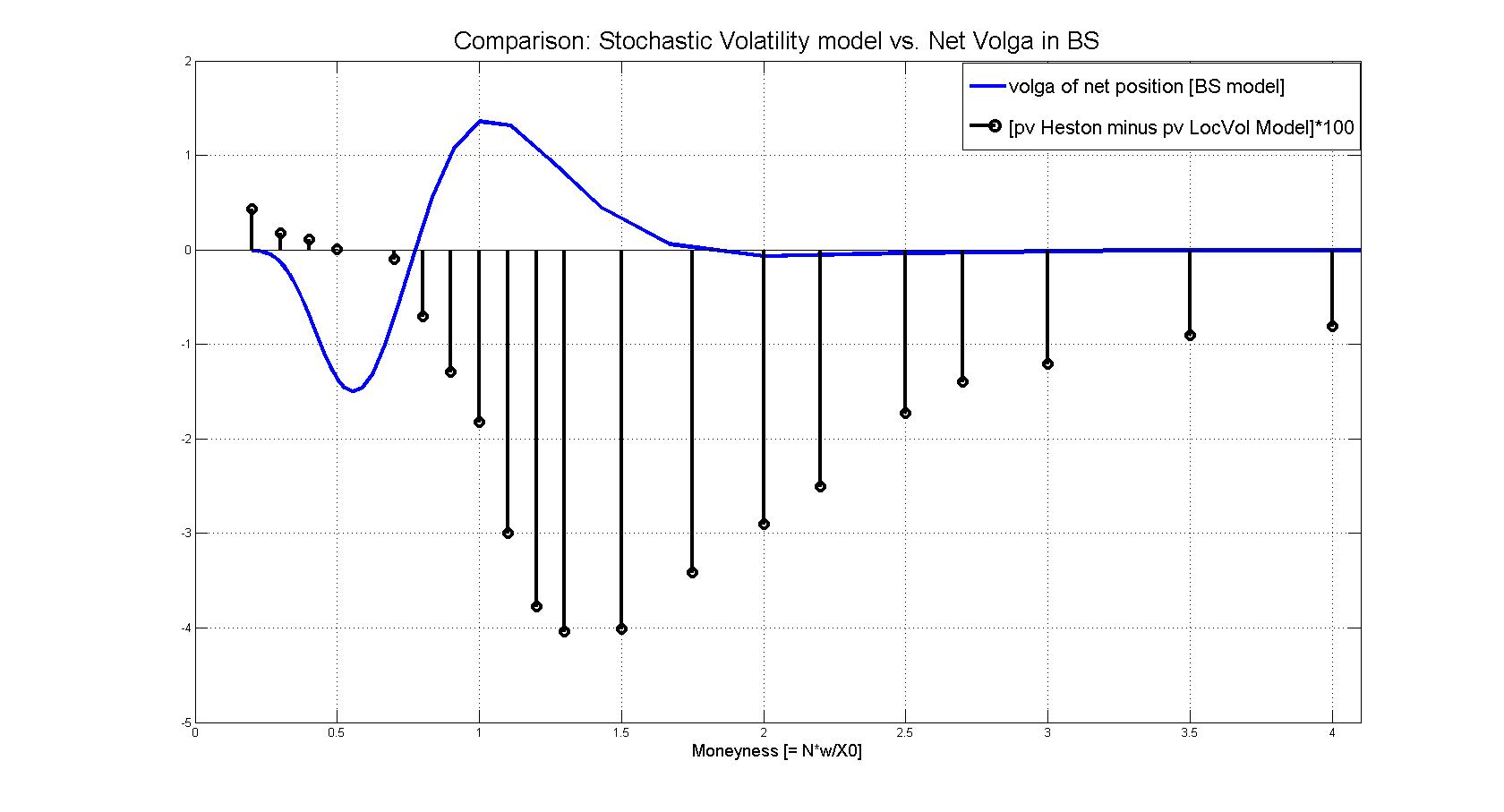}
\caption{For the two-period withdrawal guarantee,
the graph displays how 
the difference between Heston and local volatility  model value depends
on the moneyness of the guarantee and compares this dependence
with the volga of the net position after hedge in the Black-Scholes setting.
The Heston model value is below the local volatility model value mainly for those moneyness levels
that show a
long volga net position in the Black Scholes setting; vice versa for moneyness levels with short volga net position.
The Heston model parameters are as follows: initial and long-term vol $=26.5\%$, mean
reversion $=1.0$, vol-of-vol $=0.8$ and spot-vol-correlation $=0$.
The local volatility model is calibrated to fit the Heston model value of the vanilla option market.
The Black-Scholes setting for the net volga corresponds to that of figure \ref{fig:2Period_VegaVolga_BS}.}
\label{fig:2Period_Comp_HestLV_VolgaBS}
\end{center}
\end{figure}

\InsertFig{\ref{fig:2Period_Comp_HestLV_VolgaBS}}

Figure \ref{fig:2Period_Comp_HestLV_VolgaBS} shows the model value
differences of the Heston model and the corresponding local volatility
model\myfootnote{The local
volatility model is calibrated to fit the Heston model values of the
vanilla option market.} 
and compares these differences to the above volga
analysis.
As expected, the region of moneyness levels where the Heston model values are below
the local volatility model values
coincides approximately with the long volga region for the net position after hedge
in the Black-Scholes setting; vice versa for the short volga region.

\section{Multi-Period Setting}
\label{sec:MultiPeriodSetting}

We extend the semi-static hedging hedging techniques 
to the multi-period
case with guaranteed withdrawals at times $T_1, \dots, T_N$.

\subsubsection*{General One-Factor Markov Case }

Assuming (A1), we analyze 
the costs ${V}_t$ for settling the claim $Z^{(t)}$  at time $T_t$
and setting up the hedging costs for claims occurring later than
$T_t$.\myfootnote{Recall the definition of $V_t$ in
(\ref{eqn:defGuaranteeNPVPorcessVt}).}

Applying the tower law for conditional expectation to definition
\eqref{eqn:defGuaranteeNPVPorcessVt} of $V_t$, we obtain the
 recursion relation
\begin{eqnarray}
\label{eqn:GenericRecursionVt} V_t = Z^{(t)} + \E_t^\Q[V_{t+1}]
\qquad (t = 0, \dots, N-1) \, .
\end{eqnarray}
Together with the recursion \eqref{eqn:FundValueRecurrion} of the
fund value $X_t$, we deduce that the process $V$ is Markovian and
that $V_t$ depends only on $X_t$ and $S_t$ or, equivalently, on
$\zeta_{t-1}$ and $S_t$, i.e. $V_t = V_t(S_t \, | \, \zeta_{t-1})$.
In addition, the propagation from $V_t$ to $V_{t+1}$ is fully
specified by the one-period return $S_{t+1}/S_t$
of the underlying. Hence,
if the fund does not deplete at time $T_t$, i.e. $\tau >t$, the
conditional expectation in \eqref{eqn:GenericRecursionVt} can be
written as 
\begin{eqnarray}
\label{eqn:GenericRecursionVt_IntegrRepres} 
\E_t^\Q[V_{t+1}] = \int_0^\infty V_{t+1}(k \, |\,
\zeta_t = \zeta_{t-1}-w/S_t) \, \Q(S_{t+1}\in dk | S_t) \, .
\end{eqnarray}

If we now condition separately on the events $\{\tau <t\}$, $\{\tau
=t\}$, and $\{\tau >t\}$ as in \eqref{eqn:2Period_V1} and apply the
same static hedging technique\footnote{Identify
    $\Q(S_{t+1}\in dk |
    S_t) = \partial_\Kcal^2 P_t(k|S_t)$ and
    perform twice integration by parts.}
to \eqref{eqn:GenericRecursionVt_IntegrRepres} 
as in the two-period case, we arrive at the
following backwards relation: for $t=0, \dots, N-2$,
\begin{eqnarray}
\label{eqn:MuliPeriod_Tilde_Vt}
    \lefteqn{{V}_t(S_t \, | \,  \zeta_{t-1} ) = \one_{\tau < t} \cdot (N-t+1)w }\\
     &&+\one_{S_t \leq {w}/{\zeta_{t-1}}} \cdot \Big[  (N-t+1)w - \zeta_{t-1} S_t \Big]\nonumber\\
     &&+\one_{S_t >    {w}/{\zeta_{t-1}}} \cdot \int_{\frac{w}{\zeta_{t-1}-\frac{w}{S_t}}}^\infty
        P_t(k\, | \, S_t) \cdot \left[D_{S_{t+1}}^2 {V}_{t+1} \left(
           S_{t+1} \, \big| \,  \zeta_{t-1}-\frac{w}{S_t} \right) \right]_{{S_{t+1}=k}} dk \nonumber
        \nonumber
\end{eqnarray}
with terminal condition (equivalent to relation
(\ref{eqn:2Period_V1}) for the two-period case)
\begin{eqnarray*}
{V}_{N-1}(S_{N-1} \, | \,  \zeta_{N-2})
    = \mbox{$\left(\zeta_{N-2} - \frac{w}{S_{N-1}}\right)$} \cdot
    P_{N-1}\Big(\frac{w}{\zeta_{N-2}-\frac{w}{S_{N-1}}} \Big| S_{N-1}\Big)
     \quad \mbox{if $S_{N-1} > \frac{w}{\zeta_{N-2}}$,}
\end{eqnarray*}
provided that for every $t=0, \dots, N-2$ the functions $S_{t+1}
\mapsto {V}_{t+1} \left( S_{t+1} \, \big| \,  \zeta_{t} \right)$
satisfy the smoothness and boundary conditions analogous to property
(\ref{eqn:2Period_TechCondit_BdTermVanish}) for the two-period case.
The value $V_0$ of the withdrawal guarantee at time $T_0$ would then be given by
$V_0 =  \int_{w / \zeta_0}^\infty  P_0(k) \left[ D_{S_{1}}^2 {V}_{1}  \right]_{{S_{1}=k}} dk$.

However, it is  not straightforward to work out general conditions
on the (local) volatility structure that guarantee the required
smoothness and boundary properties analogous to
(\ref{eqn:2Period_TechCondit_BdTermVanish}), since these
requirements are non-local and re-apply for every iteration step in
a different manner.
In practice however, the smoothness and boundary conditions can be
verified by direct computation for every backward iteration step. It
turns out that these conditions are satisfied for many observed
implied 
volatility surfaces with not too pronounced
smile behavior.

\subsubsection*{One-factor Exponential L\'evy Case} 

We show here that the 
scaling property (A2) simplifies considerably the expressions for the weight functions.
In addition, (A2) implies the smoothness and boundary conditions analogous to
(\ref{eqn:2Period_TechCondit_BdTermVanish}) for every time step
$T_t$.

Let us define backwards recursively the sequence of weight functions
${g}_t = g_t(k)$, that indicate the weight of put options $P_t(k)$
with strike $k$ in the hedge portfolio at withdrawal time $T_t$:
\begin{eqnarray}
\label{eqn:def_tilde_gt} {g}_t(k) := \int_w^\infty
\Gamma_{t+1}(k' \, | \, k-w) \cdot {g}_{t+1}(k') \, dk' \qquad
(t = 0, \dots, N-3) \, ,
\end{eqnarray}
with terminal weight function ${g}_{N-2}(k) := \Gamma_{N-1}(w \, |
\, k-w)$. We already encountered this terminal weight function in
the two-period case, see \eqref{eqn:2Period_BS_g}.
Note that we could have started the backward recursive definition of the weights 
even one time step later by setting the terminal weight  as
${g}_{N-1}(k)dk := \delta_w$, where $\delta_w$ is Dirac's point
measure concentrated at $\{w\}$.

We state the following \emph{result}:  assume that $S$ satisfies
(A1-2). Then for any $t = 0, \dots, N-2$ with
$\tau > t$, the hedge portfolio $V_t$ in
(\ref{eqn:MuliPeriod_Tilde_Vt}) reads
\begin{eqnarray}
\label{eqn:Vt_under_LevyAssumption} V_t =
\int_{{w}/{\zeta_{t}}}^\infty \zeta_t^2 \cdot {g}_t(\zeta_t k') \cdot
P_t(k'\, | \, S_t) \, dk' = \int_{w}^\infty  {g}_t(k) \cdot P_t(k\,
| \, X_t) \, dk \, , 
\end{eqnarray}

and 
the value $V_0$ of
the withdrawal guarantee at time $T_0$ is given by $V_0 =
\int_{w}^\infty {g}_0(k) \cdot P_t(k\, | \, X_0) \, dk$.

The proof of (\ref{eqn:Vt_under_LevyAssumption}) can be motivated as
follows: assume that the fund is not depleted at time $T_{t-1}$ and
the guarantor has purchased the portfolio of put options $P_{t-1}(k
| X_{t-1})$ composed according to (\ref{eqn:Vt_under_LevyAssumption}).
 Then each put in the portfolio expires
at time $T_{t}$ and pays out \mbox{$(k-\mbox{$X_{t-1-}\cdot\frac{S_t}{S_{t-1}}$})^+$} 
$= (k-\zeta_{t-1} S_{t})^+$. Hence the
value of the expired hedge portfolio at time $T_t$ is
\begin{eqnarray}
\label{eqn:Value_ExpirHP_Vt_minus} 
V_{t-}:=\int_{w}^\infty {g}_t(k)
\cdot (k-\zeta_{t-1} S_{t})^+ \, dk \, .
\end{eqnarray}
It can be shown by iterative application of partial integration that
\begin{eqnarray}
\label{eqn:Vt_minus_equal_Vt} V_{t-} = {V}_t\,,
\end{eqnarray}
i.e. the expired hedge portfolio can finance the claim $Z^{(t)}$ at
time $T_t$ plus the hedging costs for the claims later than $T_t$.
Details have been transferred to the appendix.

\subsubsection*{Weight Function and Adjoint Multi-Contribution Fund Process}
\begin{figure}[!h]
\begin{center}
\includegraphics[width=0.8\textwidth,height = !]{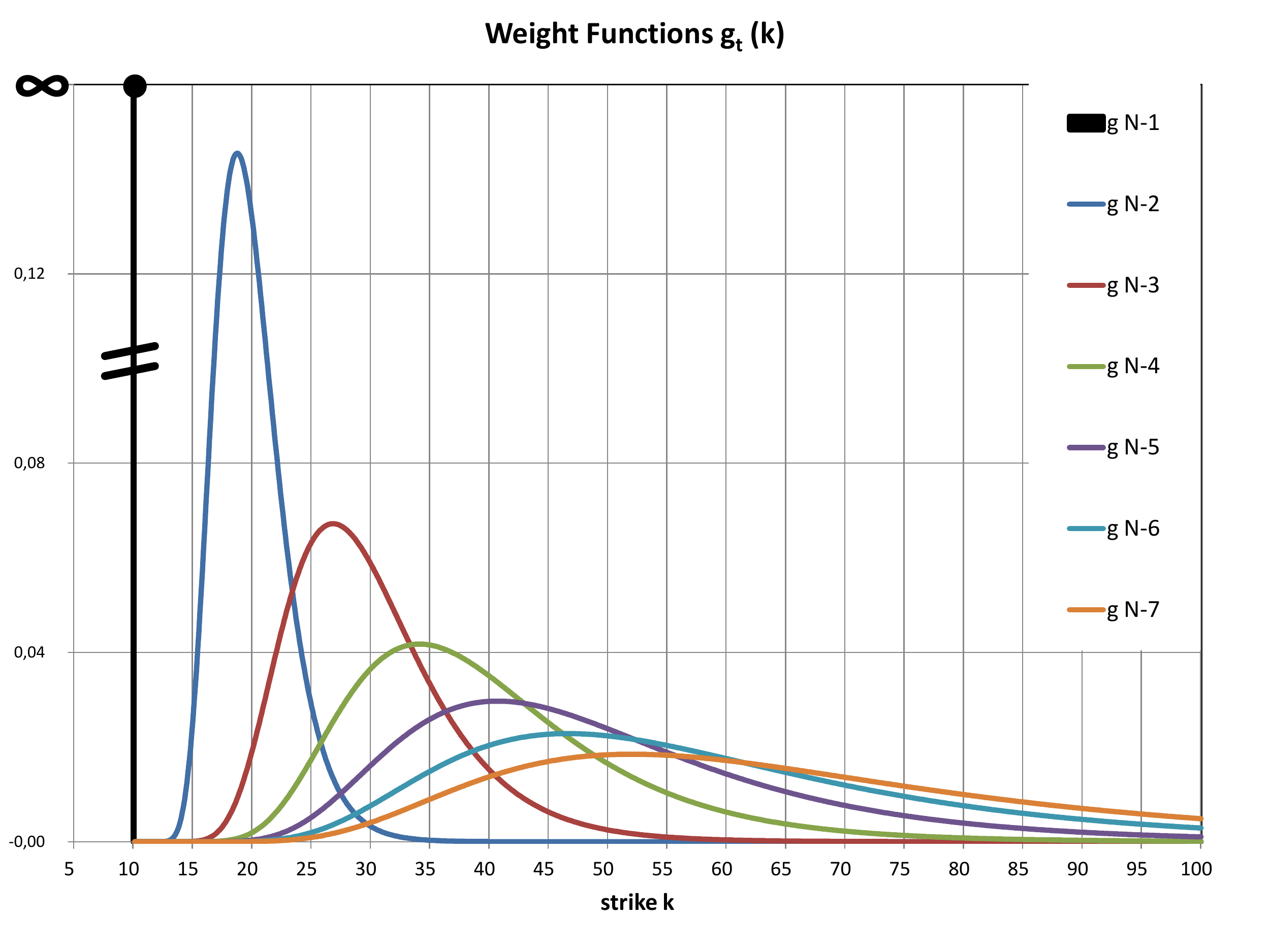}
\caption{This graphic displays the densities of the weight functions
$ {g}_t$ defined in \eqref{eqn:def_tilde_gt}, with $t$ stepping
backwards from $N-1$. The terminal weight function ${g}_{N-1}$ is
given by Dirac's  point measure concentrated at the withdrawal
amount $w=10$.
Recall that $g_t(k)dk$
describes the weight of put options in the hedge portfolio at
withdrawal time $T_t$ with strikes in a small range $dk$ around $k$.
Setting: Black-Scholes model with volatility
$=0.3$, zero interest rate and dividend yield. }
\label{fig:MeasuresGt_BlackScholes}
\end{center}
\end{figure}

The weight functions $ {g}_t$ defined in \eqref{eqn:def_tilde_gt} 
are plotted in figure \ref{fig:MeasuresGt_BlackScholes} with $t$
stepping backwards from the terminal weight function ${g}_{N-1}$
that is fully concentrated at $\{w\}$. With each time step
backwards, the weight functions are  spread wider and their mean
increases approximately by $w$. Hence the range of the strikes of
the put options in the hedge portfolios becomes wider for each
iteration and more and more put options with higher strikes are
involved.

\InsertFig{\ref{fig:MeasuresGt_BlackScholes}}

The weight functions $(g_t)_{t = N-1, \dots, 0}$
themselves represent a stochastic 
process that evolves backwards in time. We will construct
a process $(Y_t)_{t=0, \dots N-1}$ on some probability space with
measure $\tilde{\Q}$ that satisfies
\begin{equation}
\label{eqn:def_gDrivingProccY}
\tilde{\Q}(Y_t \in dk) :=
g_{N-t-1}(k) \, dk \qquad (t = 0, \dots, N-1) \, .
\end{equation}
This process $Y$ is determined by the following forward recursion relation
that results from rewriting\footnote{Replace
    in \eqref{eqn:def_tilde_gt} $t$  by $N-t-2$ and $k$ by
    $k-w$ and note that $Y\in dk+w \Leftrightarrow Y-w \in dk$.
    }
the backwards regression  \eqref{eqn:def_tilde_gt} for the weight
functions $g_t$ in terms of $Y$:
\begin{equation}
\label{eqn:gkt_QYt} \tilde{\Q}(Y_{t+1}-w \in dk) = \int_w^\infty
\Gamma_{N-t-1}(k' \, | \, k) \cdot \tilde{\Q}(Y_{t} \in dk') \qquad
(t = 0, \dots, N-2) \, .
\end{equation}
The term $\Gamma_{N-t-1}(k' \, | \, k)$ can be interpreted as
stepwise
transition density of the (shifted) 
process $Y$. In other words, the transition kernels of the
underlying asset price process that drives $Y$ must coincide with
the Gamma terms in \eqref{eqn:gkt_QYt}.

This inspires
the following \emph{definition}:
a positive valued asset price process $(\tilde{S}_t)_{t = 0, \dots,
N}$ on $\tilde{\Q}$ is called \emph{reverse $\Gamma$-adjoint} to
$({S}_t)_{t = 0, \dots, N}$ (on $\Q$), if for any time pair  $t<t'
\in \{0, \dots, N\}$ 
and $k,k'>0$
\begin{equation}
\label{eqn:DefRevAdjointGammaProc} \tilde{\Q}(\tilde{S}_{t'} \in k'+dk'
\, | \, \tilde{S}_{t} = k)  = \Gamma_{N-t',N-t}\big(k \, \big| \,
S_{N-t'}=k'\big) \, dk' \, ,
\end{equation}
where the Gamma expressions stem from vanilla options on $S$.

The following lemma summarizes some properties of 
reverse $\Gamma$-adjoint processes (the proof is transferred to the
appendix).
\begin{lemma}
\label{lemma:PropRevGammaAdjoint}
\renewcommand{\labelenumi}{\alph{enumi})}
\begin{enumerate}
\item The Gamma terms in \eqref{eqn:DefRevAdjointGammaProc}
define probability transition kernels\footnote{
    i.e. are positive as a
    function of $k\in(0,\infty)$ and integrate to one.}
and satisfy the
Chapman-Kolmogorov property. Hence \eqref{eqn:DefRevAdjointGammaProc} 
is well
defined.
\item Relation \eqref{eqn:DefRevAdjointGammaProc} allows to construct
a time-continuous version 
of a reverse $\Gamma$-adjoint process.
%
\end{enumerate}
\noindent Let $\tilde{S}$ (on $\tilde{\Q}$) be reverse
$\Gamma$-adjoint to $S$ (on $\Q$).
\begin{enumerate}
\setcounter{enumi}{2}
\item  If $S$ is a $\Q$-martingale\footnote{This
    is the case here since zero interest rates are assumed for simplicity.},
  then $\tilde{S}$ is a $\tilde{\Q}$-martingale. \label{lemma:PropRevGammaAdjoint_Martingale}
\item If $S$ satisfies (A2), so does $\tilde{S}$ and its log-return density is given by\footnote{
  In case of non-zero interest rates, the right hand side \eqref{eqn:LemRevGamAdjDensity}
  is multiplied by the discount factor $DF$. 
}
\begin{equation}
\label{eqn:LemRevGamAdjDensity}
\tilde{\Q}\left(\ln\mbox{$\frac{\tilde{S}_{t'}}{\tilde{S}_{t}}$} \in \xi + d\xi\right)
= e^{-\xi}\cdot{\Q}\left(\ln\mbox{$\frac{{S}_{N-t'}}{{S}_{N-t}}$} \in -\xi + d\xi\right)
\qquad (t<t', \, \xi\in\R)\, .
\end{equation}
\item If $S$ satisfies (A2) and \emph{Put-Call-Symmetry}\footnote{This
    is equivalent to assuming that the distribution of
     $\log(S_t)$
    is symmetric for every $t>0$,
    see e.g. Carr et al. \cite{CarrStaticHedgingExoticOption}.
    },
\eqref{eqn:DefRevAdjointGammaProc} reduces to
$\frac{\tilde{S}_{t'}}{\tilde{S}_{t}} \stackrel{d}{=}
\frac{{S}_{N-t}}{{S}_{N-t'}}$ for every $t<t'$.
\footnote{Here, $\stackrel{d}{=}$ denotes equality in
    distribution.
    }
In particular, a Black-Scholes process with
time-independent volatility is its own reverse $\Gamma$-adjoint
process.
\end{enumerate}
\end{lemma}

\begin{figure}[!h]
\begin{center}
\includegraphics[width=1.0\textwidth,height = !]{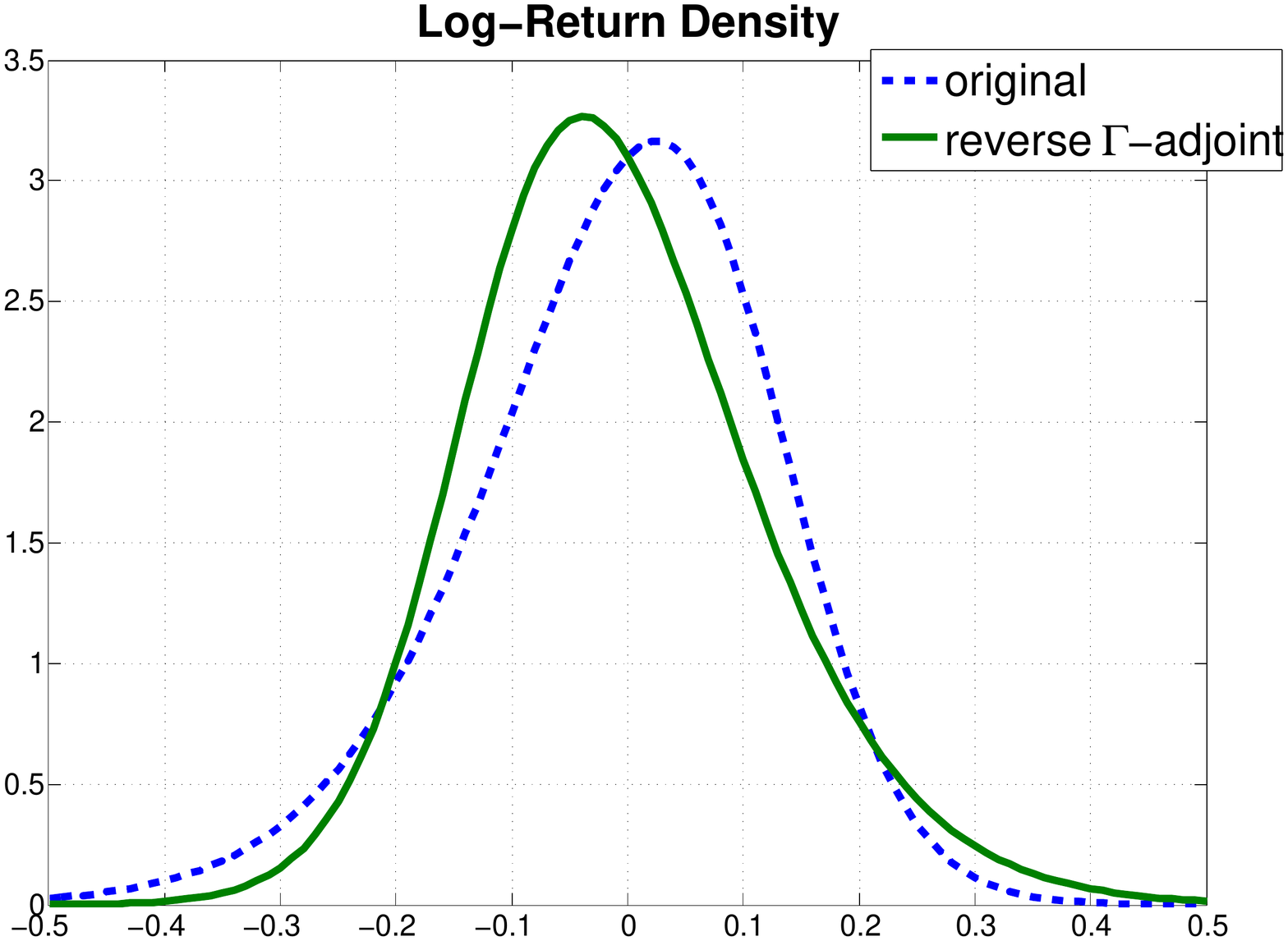}
\caption{
Density of the log-return $\ln(S_t/S_0)$ of the original process $S$ and the reverse
$\Gamma$-adjoint process in the variance gamma model.
The parameter are set to those reported in the seminal paper \cite{MadanCarrChangVarianceGamma}, Table II, 
on the variance gamma model. Interest rates are set to $2\%$ flat.}
\label{fig:VarianceGammaDensities}
\end{center}
\end{figure}

If the log-return
of $S$ has a left-skewed density, then the log-return of the 
reverse $\Gamma$-adjoint process $\tilde{S}$ has a
right-skewed density. 
Figure \ref{fig:VarianceGammaDensities} displays this effect for the variance gamma model\footnote{
    See \cite{MadanCarrChangVarianceGamma} for the seminal paper on
    the variance gamma model.
    }.

\InsertFig{\ref{fig:VarianceGammaDensities}}

Now choose a process  $\tilde{S}$ (on $\tilde{\Q}$) that is reverse
$\Gamma$-adjoint to $S$ (on $\Q$).
Since  \eqref{eqn:DefRevAdjointGammaProc} implies that
$\Gamma_{N-t-1}(k \, | \, k') \, dk' =\tilde{\Q}(\tilde{S}_{t+1}\in
dk' \, | \, \tilde{S}_{t} = k)$, 
the recursion \eqref{eqn:gkt_QYt} of the process $Y$ has the
following equivalent representation in  terms of the reverse
$\Gamma$-adjoint process $\tilde{S}$:
$$Y_{t+1} - w = Y_t \cdot {\tilde{S}_{t+1}}/{\tilde{S}_{t}}
 \qquad (t= 0, \dots, N-2) \, .$$
Since this recursion coincides with that in
\eqref{eqn:DefRecursionMultiContribFund}, 
the process $Y$ satisfying \eqref{eqn:def_gDrivingProccY} is given by a
multi-contribution fund driven by $\tilde{S}$ with regular
contribution $w$.

The mean of the multi-contribution fund value $Y_t$ is easily
obtained by relation \eqref{eqn:MultiContribFundFormula}:
$$\E^{\tilde{\Q}}[Y_t]
= w \cdot \E^{\tilde{\Q}}\left[ \sum_{u=0}^t
\frac{\tilde{S}_t}{\tilde{S}_u}\right] = w \cdot
\E^{\tilde{\Q}}\left[ \sum_{u=0}^t
\E^{\tilde{\Q}}_u\left[\frac{\tilde{S}_t}{\tilde{S}_u}\right]\right]
= w \cdot \E^{\tilde{\Q}}\left[ \sum_{u=0}^t
\frac{\tilde{S}_u}{\tilde{S}_u}\right] = w \cdot (t+1) \, ,$$
using the martingale property of the reverse $\Gamma$-adjoint
process $\tilde{S}$ (due to lemma \ref{lemma:PropRevGammaAdjoint})
and the tower law for conditional expectation.

Hence we have shown the following \emph{results}:
let $\tilde{S}$ (on $\tilde{\Q}$) be reverse $\Gamma$-adjoint to $S$
and let   $(Y_t)_{t = 0, \dots, N}$  be the multi-contribution fund
on $\tilde{S}$ according to \eqref{eqn:DefRecursionMultiContribFund}
with regular contribution $w$. Then for $t = 0, \dots, N-1$, the
weight function $g_t$ defined in \eqref{eqn:def_tilde_gt}
is generated by the probability density  of $Y_{N-t-1}$, i.e.
relation \eqref{eqn:def_gDrivingProccY} holds.
Since $Y_t \geq w$ for every $t$ by construction, the weight
function $g_t$  represents a probability density on $[w,\infty)$ with mean
\begin{eqnarray}
\label{eqn:mean_weight_Gt}
m_t &:=& \int_w^\infty k \cdot {g}_t(k) \, dk =
\E^{\tilde{\Q}}[Y_{N-t-1}] = (N-t)\cdot w \qquad 
(t = 0, \dots, N-1) \, ,
\end{eqnarray}
which proves the intuition from figure
\ref{fig:MeasuresGt_BlackScholes}.

Assuming (A1-2), the  hedging portfolio $V_t$ in the static hedging
representation \eqref{eqn:Vt_under_LevyAssumption} for the
withdrawal guarantee at time $T_t$ with $\tau > t$ (which coincides
with the withdrawal guarantee value) conditional on the fund value
$X_t$ reads
\begin{equation}
\label{eqn:Vt_ExpoLevy_Y_Formulation}
V_t  = \int_w^\infty P_t(k \, | \, X_t ) \, \tilde{\Q}(Y_{N-t-1} \in
dk) = \E^{\tilde{\Q}} \Big[ P_t(Y_{N-t-1} \, | \, X_t ) \Big] \qquad
(t = 0, \dots, N-1) \,.
\end{equation}
In the continuous time limit
\eqref{eqn:ContinTimeLimit}, this relation reads
$$V_t =
\E^{\tilde{\Q}} \big[ (\bar{Y}_{T^*-t} -\bar{X}_t)^+ \big] \qquad
\left(t\in (0,T^*] \, \right) \, ,$$ i.e. the withdrawal guarantee
is a call option on the reverse adjoint multi-contribution fund.

Relation \eqref{eqn:Vt_ExpoLevy_Y_Formulation} allows 
to calculate the value of withdrawals guarantees in a very flexibel manner,
that is particularly efficient for sensitivity within the risk management process.
We demonstrate the flexibility of this technique by applying it to the variance gamma
model.\footnote{Using 
	the calibration from figure \ref{fig:VarianceGammaDensities}.} 
The density of the log-return of
the underlying asset price process $S$ is known.\footnote{See 
	equation (23) in \cite{MadanCarrChangVarianceGamma}.
	} 
Hence the corresponding log-return
density of the reverse $\Gamma$-adjoint asset process $\tilde{S}$ is
 given by lemma \ref{lemma:PropRevGammaAdjoint}.(d). Sampling
from this density, we obtain a discrete 
approximation of the distribution of the associated
multi-contribution fund $Y$ on $\tilde{S}$ at time $T_{N-1}$.
The value of the portfolio
\eqref{eqn:Vt_ExpoLevy_Y_Formulation} of short-term put options is then
easily calculated either using the implied volatility skew curve of the
fitted variance gamma model (which is a generic outcome of the
calibration process) or using directly the market implied volatilities.

\begin{figure}[!h]
\begin{center}
\includegraphics[width=1.0\textwidth,height = !]{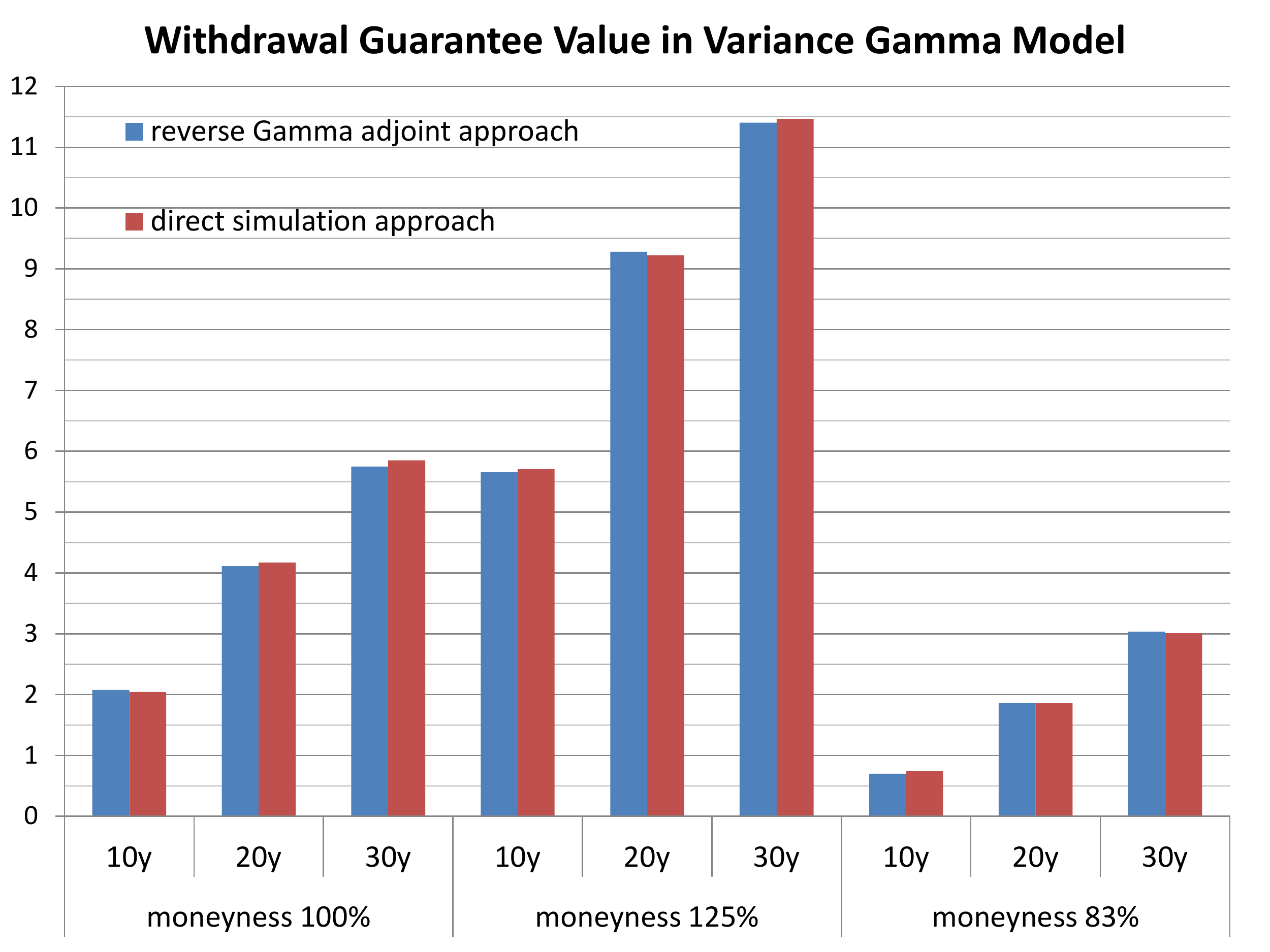}
\caption{
Net present value of withdrawal guarantees in variance gamma model
for maturities of 10, 20 and 30 years and different moneyness levels:
the blue values are obtained by sampling from the 
density of the reverse $\Gamma$-adjoint asset process 
and calculating the density of the associated multi-contribution fund 
according to \eqref{eqn:Vt_ExpoLevy_Y_Formulation}.
These results are compared with a straightforward 
Monte-Carlo valuation (red) that samples directly from the distribution of
the underlying. The variance gamma model is calibrated as in figure \ref{fig:VarianceGammaDensities}, 
regular withdrawals are assumed to occur quarterly with unit amount and moneyness
is defined as total sum of withdrawal amounts in percent of initial fund value.
Monte-Carlo simulations are performed with $10^5$ scenarios.}
\label{fig:VarianceGammaCompResults}
\end{center}
\end{figure}

Figure \ref{fig:VarianceGammaCompResults} compares the values
of withdrawal guarantees derived from this approach with straightforward 
Monte-Carlo pricing by directly sampling from the log-return distribution of
the underlying. The results nearly coincide for
different maturities and moneyness levels. The computation time for sampling 
the density of $Y_{N-1}$ is comparable to the
direct Monte-Carlo approach.\footnote{Ca.
    0.8 seconds for a 20 year
    guarantee with quarterly withdrawals based on $10^5$ scenarios using
    Matlab.}
The calculation of the value of short-term put option portfolio is
quasi immediate.\footnote{Less 
	than 0.01 seconds using a discrete
    approximation of the density $Y_{N-1}$ into 50 buckets.}
Hence the calculation of sensitivities 
with respect to shocks in the fund value and short term volatility
of the underlying (assuming that the variance gamma model parameters 
remain unchanged for maturities beyond the next withdrawal time) is speeded
up significantly. 

\InsertFig{\ref{fig:VarianceGammaCompResults}}

\subsubsection*{Options on Multi-Contribution Funds}

Vanilla options on a multi-contribution fund allow for an analogous
dual semi-static hedging representation based on 
the reverse adjoint withdrawal fund process.
We present the following \emph{result} that can be shown in analogy
to the proofs of the relations \eqref{eqn:Vt_under_LevyAssumption}
and \eqref{eqn:Vt_ExpoLevy_Y_Formulation}:

Let $(Y_t)_{t = 0, \dots, N}$ be the multi-contribution fund driven
by an asset price process $(\tilde{S})_{t = 0, \dots, N}$ on
$\tilde{\Q}$ with regular contribution $p$ according to
\eqref{eqn:DefRecursionMultiContribFund}. Consider a put option on
$Y_{N-}$ with strike $K>0$ that pays out $\tilde{V}_N :=
(K-Y_{N-})^+$ at time $T_N$.\footnote{Here $Y_{N-}$ denotes the fund
before the contribution at time $T_N$, i.e. $Y_{N-} = Y_N -p$.}

Assume that $\tilde{S}$ (on $\tilde{\Q}$) satisfies (A1-2). Then the
put option allows for a perfect semi-static hedging strategy with
hedge portfolio (and hence guarantee value) at time $T_t$
conditional on the 
fund value $Y_t$ given by
$$\tilde{V}_t = \int_{0}^\infty \tilde{g}_t(k) \cdot P_t(k\, | \,
Y_t) \, dk \qquad (t = 0, \dots, N-1) \, ,$$
where the weight
functions $\tilde{g}_t$ are backwards recursively defined by
$$ \tilde{g}_t(k) := \int_0^\infty
\Gamma_{t+1}(k' \, | \, k+p) \cdot \tilde{g}_{t+1}(k') \, dk' \quad
(t = 0, \dots, N-2) \, , \qquad \tilde{g}_{N-1}dk' := \delta_K\,.$$

Choose an asset price process ${S}$ on
${\Q}$ that is reverse $\Gamma$-adjoint to $\tilde{S}$ and let $X$
be the withdrawal fund process defined in
\eqref{eqn:FundValueRecurrion} with withdrawal amount $p$ and
initial value $X_0 = K$. Then $\tilde{g}_t$ is given by the density
of the withdrawal fund $X_{N-t-1}$, i.e. $\tilde{g}_t(k)dk =
{\Q}(X_{N-t-1} \in dk)$ for $t=0, \dots, N-1$ and
$$\tilde{V}_t  = \int_0^\infty P_t(k \, | \, Y_t ) \, {\Q}(X_{N-t-1} \in dk) =
\E^{\Q} \Big[ P_t(X_{N-t-1} \, | \, Y_t ) \Big] \qquad (t = 0,
\dots, N-1) \, .$$
In the continuous time limit \eqref{eqn:ContinTimeLimit}, this
relation reads
$$\tilde{V}_t = \E^{\Q} \big[ (\bar{X}_{T^*-t}
-\bar{Y}_t)^+ \big] \qquad \left(t\in (0,T^*] \, \right) \, ,$$ i.e.
a put on the multi-contribution fund is a call option on the reverse
adjoint withdrawal fund.

\subsubsection*{Roll-Up Feature}

A roll-up or ratchet mechanism, which is a typical additional feature in
withdrawal guarantee products, increases the future guaranteed
periodic withdrawal amounts if the underlying fund outperforms in
order to lock in the profit. The recursion
\eqref{eqn:FundValueRecurrion} for the fund value process $X$
changes to
$X_{t+1} = X_t \cdot
S_{t+1}/S_t - w_t$, 
where $w_t$ is the withdrawal amount fixed
at time $T_t$.

We formulate the roll-up mechanism in a sufficiently general manner
in terms of a guarantee base function $A_t$:
at time $T_{t}$, the guaranteed withdrawal level $w_t$ is 
increased to the fraction of the
fund value $X_t$ divided by the guarantee base $\DA_{t}$, if
this fraction exceeds the previous guaranteed withdrawal level
$w_{t-1}$, i.e.
\begin{eqnarray}
\label{eqn:Ratchet_Def_wt}
w_t  
= \max\left(w_{t-1},X_t /\DA_t\right) \qquad (t=1, \dots, N-1)\,,
\end{eqnarray}
where the guarantee base is defined as  $\DA_{t} :=
\sum_{u=t+1}^N (1+r_R)^{-1}$ with roll-up rate $r_R$. The quantity
$\DA_{t}$ can be interpreted as discounted value according to the rate
$r_R$ of future
withdrawals of unit amount later than $T_t$. 
A typical roll-up rate is $r_R=0$, i.e. $\DA_{t} = N-t$. Note
that  the limiting case $r_R\searrow -1$ (and hence $\DA_{t} \nearrow \infty$) reduces to
the situation without roll-up.
There is no roll-up event at maturity $T_N$; hence we set $\DA_{N} := \infty$ so that $w_N = w_{N-1}$.

Assuming (A2),  the withdrawal guarantee value (without roll-up)
$V_t$ at time $T_t$ depends only on the fund value $X_t$ at time
$T_t$, see relation \eqref{eqn:Vt_under_LevyAssumption}. Since 
 the  withdrawal guarantee level $w_t$ 
is a function of $X_t$ and $w_{t-1}$ by
\eqref{eqn:Ratchet_Def_wt}, the value $V^R_t$ of the withdrawal
guarantee with roll-up (or its hedge portfolio) at time $T_t$  depends only on $w_{t-1}$ and $X_t$, i.e. $V^R_t =
V^R_t(w_{t-1}, X_t)$.

If we rewrite the recursion \eqref{eqn:MuliPeriod_Tilde_Vt} in terms
of $X_t =\zeta_{t-1}S_t-w_{t-1}$, 
we obtain the following backwards recursion for $V^R_t$:
\begin{eqnarray}
\label{eqn:Ratchet_BcwdsRecursion_VRt}
V^R_t(w_{t-1}, X_t) & =& \one_{\tau < t} \cdot (N-t+1) \cdot w_{t-1}
  +\one_{w_{t-1} \leq X_t < 0} \cdot \Big[  (N-t) \cdot w_{t-1} - X_t
  \Big]\\
  &&+\, \one_{X_t>0 } \cdot \bar{V}^R_t\left(w_t =
  \max(w_{t-1},\mbox{$\frac{X_t}{\DA_t}$}), X_t\right)
  \qquad (t = 0, \dots , N-1)\, ,\nonumber
\end{eqnarray}
with
\begin{eqnarray}
\label{eqn:Ratchet_Def_BarVRt}
\bar{V}^R_t(w_{t}, X_t) :=
     \E^\Q_t\left[V^R_{t+1}(w_{t},X_{t+1} ) \, \Big| \,
     X_t\right] \qquad (t = 0, \dots
    , N-1) \, .
\end{eqnarray}
The terminal condition is $V^R_N(w_{N-1}\,|\, X_N) = Z^(N) = 
w_{N-1} \cdot \one_{\tau <N} + X_N^- \cdot \one_{\tau = Nt}$, 
see in
\eqref{eqn:DefZt}.

We show that 
the withdrawal guarantee 
$\bar{V}^R$ with roll-up feature 
admits a semi-static hedging representation (which is not 
surprising since roll-up guarantees are known to be
semi-statically hedgeable, see e.g.
\cite{CarrStaticHedgingExoticOption}).

We define
the sequences of constants $\alpha_t$ and $\beta_t$ and weight
functions $g^R_t = g^R_t(k)$ for $t = 0,\dots, N-2$ backwards
recursively by

\begin{eqnarray}
\label{eqn:Ratchet_Def_AlphBetWeights_gt}
\alpha_t &:=&\alpha_{t+1}\cdot\left[ 1+ \mbox{$\frac{1}{\DA_{t+2}}$}
\cdot C_{t+1}(1+\DA_{t+2} | \DA_{t+1}) \right]-\beta_{t+1}\cdot
P_{t+1}(1+\DA_{t+2}
| \DA_{t+1}) \nonumber\\
&&+\int_1^{1+\DA_{t+2}} P_{t+1}(k| \DA_{t+1}) \cdot g^R_{t+1}(k) \, dk\, ,\nonumber\\
\nonumber\\
\beta_t &:=& -\Big[\mbox{$\frac{\alpha_{t+1}}{\DA_{t+2}}$} \cdot
\Delta^C_{t+1}(1+\DA_{t+2} | \DA_{t+1})-\beta_{t+1}\cdot
\Delta^P_{t+1}(1+\DA_{t+2}
| \DA_{t+1}) \\
&&\quad+ \int_1^{1+\DA_{t+2}} \Delta^P_{t+1}(k| \DA_{t+1}) \cdot g^R_{t+1}(k) \, dk\Big]\, ,\nonumber\\
\nonumber\\
g^R_t(k) &:=& \left(\mbox{$\frac{\alpha_{t+1}}{\DA_{t+2}}$}-
\beta_{t+1}\right)\cdot \Gamma_{t+1}(1+\DA_{t+2} |
k-1)\nonumber\\
&&+\int_1^{1+\DA_{t+2}} \Gamma_{t+1}(k'| k-1) \cdot g^R_{t+1}(k') \,
dk' \qquad \big(k\in[1,1+\DA_{t+1}]\big)\, , \nonumber
\end{eqnarray}
with terminal conditions\footnote{Recall that $\DA_{N}$ is set to
$\infty$.
  }
$$\alpha_{N-1} = \beta_{N-1} = 0 \, , \qquad g^R_{N-1}(k)dk =
\delta_1 \, ,$$
where $\delta_1$ is the Dirac measure at $\{1\}$ and
$C_{t}$ 
denotes the call analogously to the put $P_t$; 
$\Delta^C_{t}$ and $\Delta^P_{t}$ is the Delta derivative of the
call $C_t$ and the put $P_t$, respectively.

We state the following \emph{result}: assuming (A1-2), the
withdrawal guarantee with roll-up admits a semi-static hedging
representation. The hedge portfolio 
(that also determines the guarantee
value) 
at time $T_t$ with $\tau >t$ conditional on the fund
value $X_t
>0$ and the current withdrawal level $w_t$ is given by
\begin{eqnarray}
\label{eqn:Ratchet_Result_ValueVRt}
\bar{V}^R_t(w_{t}, X_t) &=&
   \alpha_t \cdot\left[ w_t + \mbox{$\frac{1}{\DA_{t+1}}$} \cdot C_{t}\Big(w_t\cdot(1+\DA_{t+1})
     \,\Big|\,  X_{t}\Big) \right] \nonumber \\
   &&+ \; \beta_{t}\cdot P_{t}\Big(w_t\cdot(1+\DA_{t+1}) \,\Big|\
     X_{t}\Big)\\
 &&+ \; \int_1^{1+\DA_{t+1}} P_{t}(w_t \cdot k\,|\ X_{t}) \cdot g^R_{t}(k) \, dk\, ,
 \qquad (t=0,\dots, N-1) \, . \nonumber
\end{eqnarray}

\emph{Remarks:} 
(a) Note that the terminal value reads $\bar{V}^R_{N-1}(w_{N-1},
X_{N-1}) = P_{N-1}(w_{N-1}|X_{N-1})$.

(b) $\bar{V}^R_t$ also satisfies (A2), i.e.
if the fund value and the withdrawal level double (and the moneyness
of the guarantee remains constant) the absolute guarantee value also
doubles.
This implies that when the fund value $X_t$ exceeds the level $\DA_t
w_{t-1}$ and the roll-up event occurs, the guarantee value is linear
in the fund value $X_t$.\footnote{The
    guarantee value then reads
    $\bar{V}^R_t(X_t/\DA_t,X_t) =X_t\cdot\bar{V}^R_t(1/\DA_t, 1)$.}

(d) Relation \eqref{eqn:Ratchet_Result_ValueVRt} is again proved 
 by applying static hedging
techniques that allow to identify
\begin{eqnarray}
\label{eqn:Ratchet_Identif_AlphBetaGRt}
  \alpha_t = \bar{V}^R_{t+1}(1,\DA_{t+1}) \,,  \;
  \beta_t  =\partial_{X_{t+1}} \bar{V}^R_{t+1}(1,\DA_{t+1}) \,,  \;
  g^R_t(k) =\partial_{X_{t+1}}^2 \bar{V}^R_{t+1}(1,k-1) \,.
\end{eqnarray}
The details of the proof are transferred to the appendix. 
Note that not only put options are used in the static hedging
representation as for the non-roll-up type 
but also call options are involved here.

(d) It can be shown that the static hedging representation
\eqref{eqn:Ratchet_Def_AlphBetWeights_gt} and
\eqref{eqn:Ratchet_Result_ValueVRt} 
converges in the limit $\DA_t
\nearrow \infty$ to the non-roll-up case \eqref{eqn:def_tilde_gt}
and \eqref{eqn:Vt_under_LevyAssumption}.\footnote{It suffices to
    show by backwards induction that $\alpha_t,\beta_t \to
    0$.
    }

\subsubsection*{Concluding Remarks and Outlook}  

\begin{figure}[!h]
\begin{center}
\includegraphics[width=1.0\textwidth, height = 0.45\textheight]{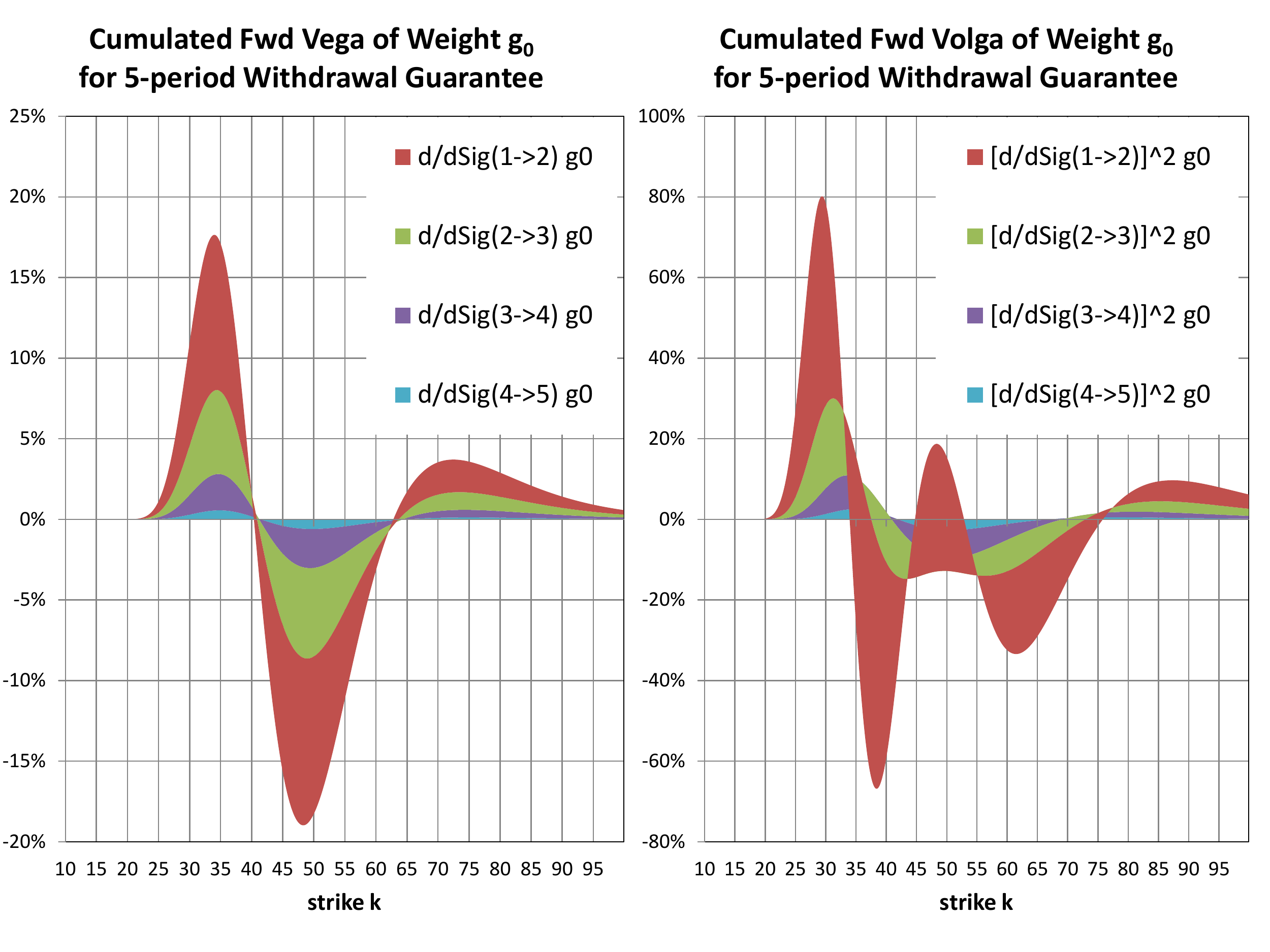}
\includegraphics[width=0.7\textwidth, height = 0.40\textheight]{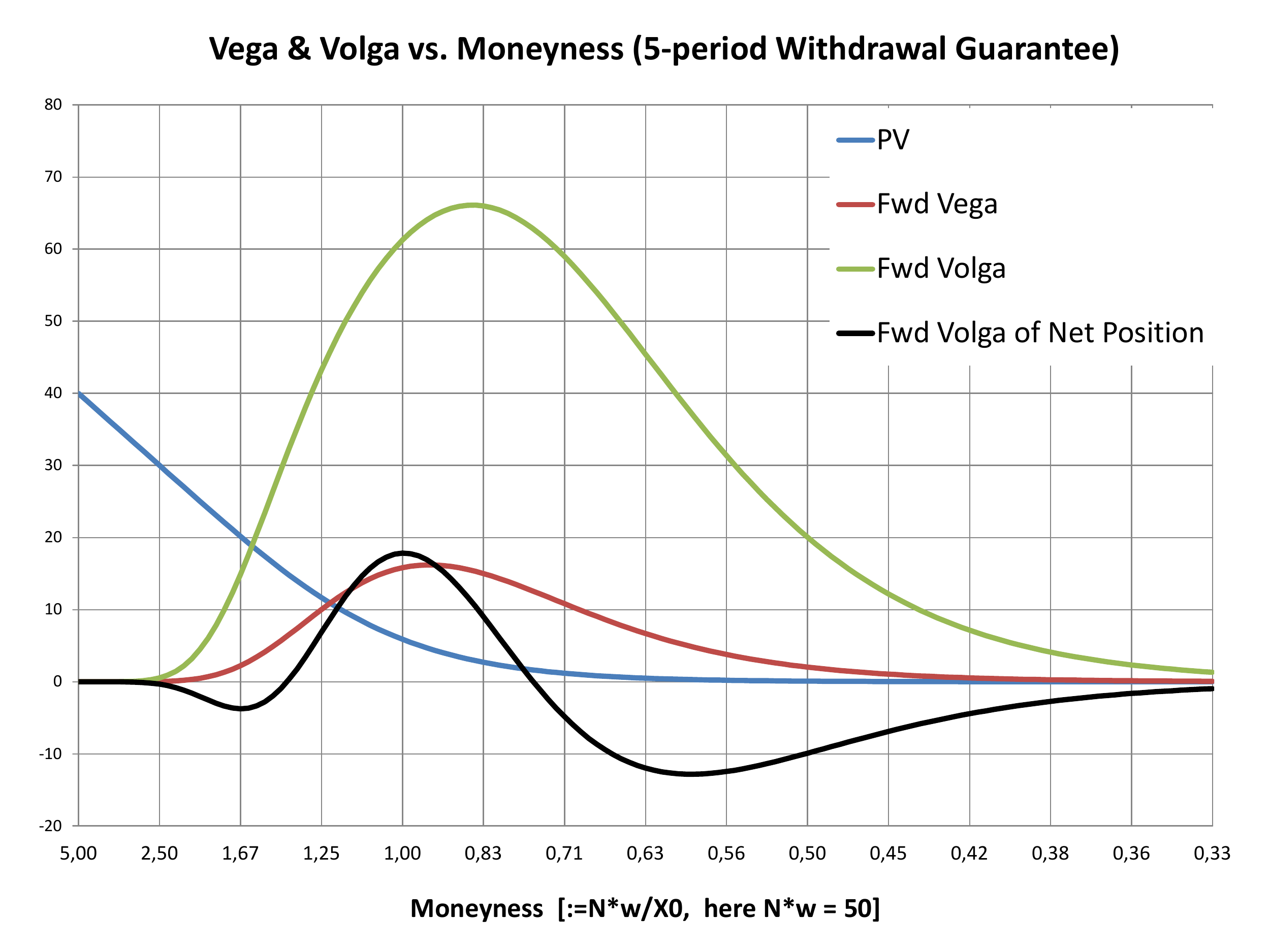}
\caption{
For a 5-period withdrawal guarantee, the top graph shows the forward vega and
volga sensitivity of the weight function $g_0$, split into the
sensitivity contributions with respect to the $T_t$-$T_{t+1}$-forward volatilities
$\Sigma_t$ for $t=1, \dots, 4$.
The bottom graph displays present value and forward vega and volga
sensitivities of the guarantee for different moneyness levels
together with the volga of the net position after hedge with forward
starting variance swaps.
The results are very similar to those in the two-period case, see figure \ref{fig:2Period_VegaVolga_BS}.
Setting: Black-Scholes model as in two-period case.}
\label{fig:MultiPeriod_VegaVolga}
\end{center}
\end{figure}

\renewcommand{\labelenumi}{\arabic{enumi})}
\begin{enumerate}
\item The effects of stochastic volatility are very similar to the
two-period case.
Figure \ref{fig:MultiPeriod_VegaVolga} displays for the 5-period
case the total forward vega and volga sensitivity of the weight
function $g_0$ and the split into the sensitivities with respect to
the series of yearly forward volatilities. 
Similar to the two-period case\footnote{Compare 
	with figure \ref{fig:2Period_VegaVolga_BS}.}, 
there is a net volga long position
(after hedging with forward starting variance swaps) except for very
in- and out-of-the-money guarantees. Hence, stochastic volatility
models will price the guarantee at lower values than the
corresponding local volatility model.
\InsertFig{\ref{fig:MultiPeriod_VegaVolga}}
\item 
 For one-factor exponential L\'evy models, the semi-static
hedging representation \eqref{eqn:Vt_ExpoLevy_Y_Formulation} allows a
very efficient calculation of the withdrawal guarantee value and its
sensitivities with respect to changes in the fund value and in the
short-term volatility of the underlying. 
This is essential for the risk management  of such products.
\item 
For non-zero deterministic interest rates, the static hedging representations
\eqref{eqn:MuliPeriod_Tilde_Vt} and
\eqref{eqn:Vt_under_LevyAssumption} remain valid, since the
discounting effects are already correctly reflected in the put
option values and their derivatives.

Due to the long-term nature of typical GMWB products, stochastic
interest rates seem an indispensable ingredient for the valuation
model at first glance.
Unfortunately, our approach has no straightforward generalization to stochastic
interest rates.
Table \ref{tab:HWSensie} summarizes the interest rate volatility 
sensitivity for at-the-money GMWB products valuated in the
 one-factor Hull-White \& Black-Scholes hybrid model (using Monte-Carlo 
 techniques) and compares it to the sensitivities of vanilla put options  
und puts on multi-contribution funds: 
interest rate volatility sensitivity for withdrawal
guarantees increases with maturity to ca. 1.5\% of the 
base guarantee value for maturities of 30 years 
but remains at only ca. 30-35\% of that 
of the corresponding vanilla put with same notional.
Things are different for puts on multi-contribution funds which 
show much higher interest rate volatility sensitivity.
The reason is that the withdrawals diminish the fund value
and hence the volatility exposure relatively quickly in average 
whereas the multi-contribution fund accumulates the 
volatility exposure over run time.
\InsertTable{\ref{tab:HWSensie}}
This analysis shows that the need for stochastic interest rates for the
valuation model of GMWB products is less pronounced compared to  
long-term vanilla options and puts on multi-contribution funds.

\item The presented semi-static hedging results do not only apply to
 simple withdrawal guarantee structures introduced in section
\ref{sec:DesciptionGuarantee}. They immediately extend to
non-constant deterministic schedules of withdrawal amounts.
The technique also extends to roll-up or ratchet features, that are very common
for GMWB products, 
see \eqref{eqn:Ratchet_Result_ValueVRt}. 

\end{enumerate}

\begin{appendix}

%

\section{Appendix: Proofs}

\noindent {\bf Proof of
(\ref{eqn:2Period_TechCondit_BdTermVanish}).}
For the function $S_1 \mapsto V_1(S_1)$ that is piecewise defined
in (\ref{eqn:2Period_V1}), we show that $V_1$ and $V_1'$ is
continuous at $S_1 = w/\zeta_0$ and that $V_1' = o(1/S_1)$ as
$S_1\nearrow\infty$. The second boundary condition $V_1(0) = 2w$ is
obvious.

First note that $V_1'=  [\partial_{S_1}+(\partial_{S_1}
\zeta_1)\cdot \partial_{\zeta_1} ] \cdot[\zeta_1 \cdot P_1\left(
w/\zeta_1\right)]$ for $S_1>w/\zeta_0$ and hence
\begin{eqnarray}
\label{eqn:proof_calc_V1prime} V_1'
&=& \zeta_1 \cdot\Delta_1\left( \frac{w}{\zeta_1} \, \Big| \, S_1 \right)
+ \frac{w}{S_1^2}\cdot\left[ P_1\left(\frac{w}{\zeta_1} \,
\Big| \, S_1\right) -
\partial_\Kcal P_1\left(\frac{w}{\zeta_1} \, \Big| \, S_1 \right)\cdot\frac{w}{\zeta_1}\right]\, .
\end{eqnarray}
From the definition of $V_1$ in \eqref{eqn:2Period_V1}, we obtain
$\lim_{S_1 \nearrow w/\zeta_0} V_1(S_1) = w$ and
\mbox{$\lim_{S_1 \nearrow w/\zeta_0} V_1'(S_1) = -\zeta_0$}.
Further, $S_1 \searrow w/\zeta_0$ implies $\zeta_1 \searrow 0$.
Hence,
$$\lim_{S_1 \searrow w/\zeta_0} V_1(S_1) =
\lim_{\zeta_1\searrow 0} \zeta_1 P_1\left(w/\zeta_1 \right) =
\lim_{\zeta_1\searrow 0} \E^\Q_1 (w-\zeta_1 S_2)^+ = w \, ,$$ where
the last equality follows from monotone convergence.
Similarly, we obtain using (\ref{eqn:proof_calc_V1prime})
\begin{eqnarray*}
\lim_{S_1 \searrow w/\zeta_0} V_1'(S_1)
&=&
\lim_{\zeta_1\searrow 0} \zeta_1 \cdot\Delta_1\left( \frac{w}{\zeta_1} \, \Big| \, \frac{w}{\zeta_0} \right)
+\frac{\zeta_0^2}{w}\cdot
\lim_{\zeta_1\searrow 0} \left[ P_1\left(\frac{w}{\zeta_1} \, \Big|
\,  \frac{w}{\zeta_0}\right) -\frac{w}{\zeta_1}\cdot\partial_\Kcal
P_1\left(\frac{w}{\zeta_1} \, \Big| \,
\frac{w}{\zeta_0}\right)\right ]\\
&=&\frac{\zeta_0^2}{w}\cdot \lim_{\zeta_1\searrow 0} \E^\Q_1 \left[
\left(\frac{w}{\zeta_1} -S_2\right)\cdot\one_{{w}/{\zeta_1} >S_2}
-\frac{w}{\zeta_1}\cdot\one_{{w}/{\zeta_1} >S_2} \, \Big| \, S_1
=\frac{w}{\zeta_0}\right ]\\
&=&\frac{\zeta_0^2}{w}\cdot \lim_{\zeta_1\searrow 0} \E^\Q_1 \left[
 -S_2 \cdot\one_{{w}/{\zeta_1} >S_2} \, \Big| \, S_1
=\frac{w}{\zeta_0}\right ]\\
&=&\frac{\zeta_0^2}{w}\cdot \E^\Q_1 \left[
 -S_2  \, \Big| \, S_1
=\frac{w}{\zeta_0}\right ] =\frac{\zeta_0^2}{w}\cdot \left[
 -\frac{w}{\zeta_0}\right ] = -\zeta_0 \, ,
\end{eqnarray*}
where the second line uses the fact that $\Delta_1$ is bounded and
the first equality in the last line above follows again from
monotone convergence. This proves the continuity of $V_1$ and $V_1'$
at $S_1 = w/\zeta_0$.

Finally, $S_1 \nearrow \infty$ implies $\zeta_1 \to \zeta_0$. Hence
we obtain using (\ref{eqn:proof_calc_V1prime})
\begin{eqnarray*}
\lim_{S_1 \nearrow \infty} S_1 \cdot V_1'(S_1) &=&
\frac{w}{S_1}\cdot\left[ P_1\left(\frac{w}{\zeta_0} \, \Big| \,
S_1\right) -
\partial_\Kcal P_1\left(\frac{w}{\zeta_0} \, \Big| \, S_1 \right)\cdot\frac{w}{\zeta_0}\right] = 0 \, ,
\end{eqnarray*}
since all terms in $[\cdot]$-brackets are bounded functions of
$S_1$. Hence, $V_1' = o(1/S_1)$ as $S_1\nearrow\infty$.
\qed

The following lemmata deal with implications of the assumption (A2).
\begin{lemma}
\label{lemma:EquivLevyScalinProp}
(A2) is equivalent to assume that $S$ has independent returns,
  i.e. $S_{t+1}/S_t$ is independent on $S_t$ for every $t = 1,\dots,
  N$.
\end{lemma}
\proof 
$\Leftarrow$ is obvious. To show $\Rightarrow$,
note that by static hedging theory\\ \mbox{$\Q(S_{t+1}/S_t \leq K \,
| \, S_t = s)= \Q(S_{t+1} \leq s\cdot K \, | \, S_t = s) =
\partial_\Kcal P_t(s\cdot K \, | \, s)$.}\\ It suffices to proof
that $\partial_\Kcal P_t(s\cdot K \, | \, s)$ is independent of $s$
for
every $K>0$:\\
(A2) implies $P_t(k \, | \, s') = k\cdot P_t(1 \, | \, s'/k)$. Hence
$\partial_k P_t(k \, | \, s') = P_t(1 \, | \, s'/k) - (s'/k)\cdot
\Delta_t(1 \, | \, s'/k)$. Thus $\partial_\Kcal P_t(s\cdot K \, | \,
s) = P_t(1 \, | \, 1/K) - (1/K)\cdot \Delta_t(1 \, | \, 1/K) =
\partial_\Kcal P_t(K \, | \, 1)$.\qed

\begin{lemma}
\label{lemma:Hopital}
(A2) implies that $x\cdot \Delta_t(k\, | \, x) \to 0$ and $k\cdot
\partial_\Kcal C_t(k\, | \, x) \to 0$
 as $x,k\nearrow \infty$ uniformly in $k\in
\R$ and $x\in \R$ for every $t = 0, \dots, N-1$.
\end{lemma}
\proof Assumption (A2) implies
$P_t(k\, | \, x)= x \cdot P_t(k/x\, | \, 1)$. Hence
\begin{eqnarray}
x\cdot \Delta_t(k\, | \, x) &=& x \cdot D_x [x \cdot P_t(k/x\, | \, 1)] = 
x\left[P_t(k/x\, | \, 1) - x\cdot \partial_\Kcal P_t(k/x\, | \, 1) \cdot (k/x^2)\right] \nonumber\\
&=& P_t(k\, | \, x) - k\cdot \Q(S_{t+1} \leq k/x \, | S_t = 1) \, ,
\label{eqn:LemHopitHelp}
\end{eqnarray}
where the last equation follows from (A2) and the
Breeden-Litzenberger formula (\ref{eqn:2Period_StaticHedging}). As
$x\nearrow\infty$, both terms converge to zero locally uniformly in
$k$. The proof for $k\cdot
\partial_\Kcal C_t(k\, | \, x)$ follows from analog considerations.\qed

\begin{lemma}
\label{lemma:ScalingProp_Derivaties} If a twice differentiable
function $f=f(x,y)$ satisfies  $f(\alpha x,\alpha y) = \alpha
f(x,y)$ for every $\alpha\in\R$, then $\partial_\Xcal^n f(\alpha
x,\alpha y) = \alpha^{1-n}\partial_\Xcal^n f(x,y)$ and
$\partial_\Ycal^n f(\alpha x,\alpha y) =
\alpha^{1-n}\partial_\Ycal^n f(x,y)$ for $n\in\N$.\footnote{Here,
    $\partial_\Xcal^n f$ and $\partial_\Ycal^n f$ denotes the n-th
    partial derivative of $f$
    with respect to the $X$ and $Y$ coordinate, respectively.}
\end{lemma}
\proof Follows from $\alpha \partial_x^n f( x,y) =
\partial_x^n (\alpha f( x,y)) = \partial_x^n  f( \alpha x,\alpha y) =
\alpha^n
\partial_\Xcal^l f(\alpha x,\alpha y)$.

\vspace{3mm} \noindent {\bf Proof of (\ref{eqn:Vt_minus_equal_Vt}).}
Applying partial integration to (\ref{eqn:Value_ExpirHP_Vt_minus})
yields
\begin{eqnarray*}
    V_{t-} 
    &\stackrel{(\ref{eqn:def_tilde_gt})}{=}&
        \int_{\max(w,\zeta_{t-1}S_{t})}^\infty (k - \zeta_{t-1}
S_{t})  \cdot \left(\int_w^\infty \Gamma_{t}(k'\, | \, k-w) \, {g}_{t}(k') \, dk'\right) dk \nonumber \\
    &=& \left[(k - \zeta_{t-1}S_{t})  \int_w^\infty \Delta_{t}(k'\, | \, k-w)
    \, {g}_{t}(k') \, dk' \right]_{k = \max(w,\zeta_{t-1}S_{t})}^\infty  \\
    &&  +\left\{ -\int_{\max(w,\zeta_{t-1}S_{t})}^\infty  \int_w^\infty \Delta_{t}(k'\, | \, k-w) \,
     {g}_{t}(k') \, dk' \, dk \right\}\, . \nonumber
\end{eqnarray*}
We first analyze the term in $[\cdot]$-brackets. 
The upper bound contribution for the limit $k \nearrow \infty$
vanishes due to lemma  \ref{lemma:Hopital}.
Hence the $[\cdot]$-term has only contributions from the lower
bound, which reads
\begin{eqnarray*}
[\cdot] &=&
-\one_{\zeta_{t-1}S_{t}<w} \cdot (w-\zeta_{t-1}S_{t})  \int_w^\infty (-1) \cdot {g}_{t}(k') \, dk'\\
&\stackrel{{(\ref{eqn:mean_weight_Gt})}}{=}& -X_{t}\cdot\one_{X_{t}
< 0} =
X_t^- = Z^{(t)}   \, .
\end{eqnarray*}
When integrating the $\{\cdot\}$-term, 
we obtain
\begin{eqnarray*}
\{\cdot\} &=& -\left[ \int_w^\infty P_{t}(k'\, | \, k-w) \,
     {g}_{t}(k') \, dk' \right]_{k=\max(w,\zeta_{t-1}S_{t})}^\infty \\
&=&  \int_w^\infty P_{t}\left( k'\, \Big| \, \max(0,\zeta_{t-1}S_{t}-w) \right) \,{g}_{t}(k') \, dk' \\
&=&  \one_{X_t\leq 0} \int_w^\infty k' \, {g}_{t}(k') \, dk' + \one_{X_t > 0} \int_w^\infty p_{t}(X_{t},k') \, {g}_{t}(k') \, dk'   \\
&\stackrel{{(\ref{eqn:mean_weight_Gt})}}{=}&
    \one_{\tau = t} \cdot m_t  + \one_{\tau > t} V_t\, .
\end{eqnarray*}
Hence 
$V_{t-} = Z^{(t)} + \one_{\tau = t} \cdot m_t  + \one_{\tau > t} V_t
\stackrel{{(\ref{eqn:MuliPeriod_Tilde_Vt})}}{=} {V}_t$, which proves
(\ref{eqn:Vt_minus_equal_Vt}). \qed

\vspace{3mm} \noindent {\bf Proof of Lemma
\ref{lemma:PropRevGammaAdjoint}.}
For $l=0,1$, we deduce
\begin{eqnarray}
\label{lemma:PropRevGammaAdjoint_ProofHelp} \int_0^\infty
s^l\cdot\Gamma(K|s) ds
    &=& \lim_{s^*\uparrow \infty}
    \{s^l\Delta(K|s)]_{s=0}^{s^*}
    -l\cdot [P(K,s)]_{s=0}^{s^*}\}
    = \one_{l=0}+ K\cdot\one_{l=1} \, ,
\end{eqnarray}
since $\lim_{s^*\uparrow \infty}
    s^*\Delta(K|s^*) =0$ in case $l=1$  by lemma \ref{lemma:Hopital}.

\emph{(a):} the density $\Gamma(k'|k)dk$ integrates to one on
$k\in[0,\infty)$ by \eqref{lemma:PropRevGammaAdjoint_ProofHelp}.
Chapman-Kolmogorov (CK) property for $\tilde{S}$ follows from the CK
property of the put option value process $t\mapsto P_{t,N}(K|S_t) =
\E_t[(K-S_N)^+|S_t]$ (a martingale!) and twice differentiation with
respect to $S_t$.
\emph{(b):}
    on a discrete time grid, \eqref{eqn:DefRevAdjointGammaProc}
    generates a Markov chain on $\R^+$. By halving successively the grid size,
    a consistent sequence of Markov chains is generated due to the CK property.
    This sequence of Markov chains can shown to be tight which guarantees the existence of
    a time-continuous limit process $\tilde{S}$.
\emph{(c):} $\E_t[\tilde{S}_{t'}|\tilde{S}_{t}=k] =\int_0^\infty k'
\Gamma_{N-t',N-t}(k|k')dk' = k$ by
\eqref{lemma:PropRevGammaAdjoint_ProofHelp}.
The proof of \emph{(d)} is analog to the proof of Lemma
\ref{lemma:EquivLevyScalinProp} with $\partial_\Kcal P_t$ replaced
by $\Delta_t$.
To show the relation \eqref{eqn:LemRevGamAdjDensity} for the log-return density of the 
ajoint process we assume without loss of generality that $\tilde{S}_{t} = 1$. Hence
\begin{eqnarray*}
\lefteqn{\tilde{\Q}\left(\ln\tilde{S}_{t'} \leq \xi \, | \, \tilde{S}_{t} = 1\right) =
\tilde{\Q}\left(\tilde{S}_{t'} \leq e^\xi \, | \, \tilde{S}_{t} = 1\right) 
\; \stackrel{\eqref{eqn:DefRevAdjointGammaProc}}{=}\; 
  \int_0^{e^\xi} \Gamma_{N-t',N-t}(1|k') \, dk'}&&\\
&&= \Delta^C _{N-t',N-t}(1\, | \,e^\xi)
\; \stackrel{L\ref{lemma:ScalingProp_Derivaties}}{=} \; \Delta^C _{N-t',N-t}(e^{-\xi}\, | \,1)\\
&&= C _{N-t',N-t}(e^{-\xi}\, | \,1) + e^{-\xi}\cdot\Q(S_{N-t}>e^{-\xi} \, | \, S_{N-t'} = 1) \\
&&=\int_{e^{-\xi}}^\infty s \cdot \Q(S_{N-t} \in s+ds  \, | \, S_{N-t'} = 1) 
  =\int_{e^{-\xi}}^\infty s \cdot \mbox{$\frac{1}{s}$}\cdot \Q(\ln S_{N-t} \in \ln s + ds \, | \, S_{N-t'} = 1) \, ,
\end{eqnarray*}
where the second last equation follows analogously to relation \eqref{eqn:LemHopitHelp}.
Differentiation with resprect to $\xi$ proves relation \eqref{eqn:LemRevGamAdjDensity}.
\emph{(e):} 
It sufficies to show
\begin{equation}
\label{eqn:PutCallSymmetryGammaDensityRelation}
\Q(S_{t'} \in dk' \, | \, S_{t} = k) = \Gamma_{t,t'}(k \,| \,
S_t=k)\, dk  \qquad (t<t'\in\{0,\dots, N\}, \; k,k'>0)\, .
\end{equation}
Put-Call-Symmetry implies (see e.g. \cite{CarrStaticHedgingExoticOption})
$C_t(K | S_t) /\sqrt{K} =  P_t(K' | S_t) /\sqrt{K'}$
for any pair of strikes $K,K'\geq 0$ with $K\cdot K' = S_t^2$.
Chose $K' = S_t^2/K$. Multiplying by $\sqrt{K}$  and applying call-put-parity gives
$P_t(K | S_t) + S_t - K = C_t(K | S_t)= \sqrt{K} \cdot
(\sqrt{K}/{S_t}) \cdot  P_t\left( {S_t^2}/{K} \, | S_t\right)
\stackrel{(A2)}{=} P_t\left( S_t | K\right) \, .$ Differentiating
twice with respect to  $S_t$ and setting $K = k'$ and $S_t = k$
yields \mbox{$\Gamma_{t}(k' \, | \, k) = \partial_\Kcal^2 P_{t}(k \, | \,
k')$} and \eqref{eqn:PutCallSymmetryGammaDensityRelation} follows
from the classical static hedging relation $\partial_\Kcal^2 P_{}(k
| k') = \Q(S_{t+1} \in dk \, | \, S_{t} = k')$. \qed

\vspace{3mm} \noindent {\bf Proof of
(\ref{eqn:Ratchet_Result_ValueVRt}).}
We apply static hedging techniques to the conditional expectation
$\bar{V}^R_t$ in \eqref{eqn:Ratchet_Def_BarVRt}.
(A2) implies that the density of $X_{t+1}+w_t =
X_t\cdot(S_{t+1}/S_t)$ is given by
$$\Q(X_{t+1}+w_t \in dk) =\partial_\Kcal^2 P_t(k|X_t)\,dk =\partial_\Kcal^2
C_t(k|X_t)\,dk \, .$$
Since $V^R_{t+1}=V^R_{t+1}(w_t,X_{t+1})$ defined in
\eqref{eqn:Ratchet_BcwdsRecursion_VRt} is continuous in $X_{t+1}$
and $\partial_{X_{t+1}}V^R_{t+1}$  except at the
roll-up point $X_{t+1-}+w_t = X_{t+1}=\DA_{t+1} w_{t}$, we obtain by splitting the
integration at $X_{t+1-} = (1+\DA_{t+1}) w_t$
\begin{eqnarray*}
\bar{V}^R_t(w_{t},X_t) &=&
   \left[ \int_0^{w_t(1+\DA_{t+1})}\partial_\Kcal^2 P_t(k|X_t)
 + \int_{w_t(1+\DA_{t+1})}^\infty \partial_\Kcal^2 C_t(k|X_t)\right]
 V_{t+1}^R(w_t,k-w_t)  \,dk \\
&=& \Big[  \partial_\Kcal P_t(k|X_t) \cdot
V_{t+1}^R(w_t,k-w_t)\Big]_{k=0}^{w_t(1+\DA_{t+1})}\\
  &&+ \Big[  \partial_\Kcal C_t(k|X_t) \cdot V_{t+1}^R(w_t,k-w_t)\Big]_{k=w_t(1+\DA_{t+1})}^\infty\\
&&- \Big[  P_t(k|X_t) \cdot\partial_{X_{t+1}}
V_{t+1}^R(w_t,k-w_t-0)\Big]_{k=0}^{w_t(1+\DA_{t+1})}\\
&&  - \Big[  C_t(k|X_t) \cdot\partial_{X_{t+1}}
  V_{t+1}^R(w_t,k-w_t+0)\Big]_{k=w_t(1+\DA_{t+1})}^\infty\\
&&+\left[ \int_0^{w_t(1+\DA_{t+1})}  P_t(k|X_t)
 + \int_{w_t(1+\DA_{t+1})}^\infty  C_t(k|X_t)\right] \partial_{X_{t+1}}^2
 V_{t+1}^R(w_t,k-w_t) \, dk\\
&=&V_{t+1}^R(w_t,w_t\DA_{t+1})\\
 &&+ \, C_t\left(w_t(1+\DA_{t+1}) | X_t\right) \cdot\partial_{X_{t+1}}
   V_{t+1}^R(w_t,w_t\DA_{t+1}+0)\\
 &&- \, P_t\left(w_t(1+\DA_{t+1}) | X_t\right) \cdot\partial_{X_{t+1}}
   V_{t+1}^R(w_t,w_t\DA_{t+1}-0)\\
  &&+ \int_{1}^{1+\DA_{t+1}} P_t(w_t k|X_t)\cdot \partial_{X_{t+1}}^2
  V_{t+1}^R(w_t,w_t(k-1)) \, w_t \, dk\\
&&  + \int_{1+\DA_{t+1}}^\infty   C_t(w_t k|X_t)\cdot
\partial_{X_{t+1}}^2
  V_{t+1}^R(w_t,w_t(k-1)) \, w_t \, dk \, ,
\end{eqnarray*}
where the last equality follows from the boundary conditions for
$V_t$ (and hence also for $V^R_t$) at $X_t = 0$ analog to
\eqref{eqn:2Period_TechCondit_BdTermVanish}, the relation
$\partial_\Kcal P_t(k) = \partial_\Kcal C_t(k)+1$ stemming from
put-call-parity, the fact that $\partial_{X_{t+1}}^2
V_{t+1}^R(k-w_t,w_t) \equiv 0$ on $k\in [0,w_t]$, and the change of
integration variable $k\mapsto k/w_t$, provided that
\begin{eqnarray}
\label{eqn:Ratchet_TechGrowthCondition} \lim_{k\nearrow\infty}
\partial_\Kcal C_t(k) \cdot V_{t+1}^R(w_t,k-w_t) = \lim_{k\nearrow\infty}
 C_t(k) \cdot \partial_{X_{t+1}} V_{t+1}^R(w_t,k-w_t) = 0 \, .
\end{eqnarray}

Now suppose that $\bar{V}^R_{t+1}$ satisfies (A2), i.e.
$\bar{V}^R_{t+1}(\alpha \cdot w_{t+1}, \alpha \cdot X_{t+1}) =
\alpha \cdot \bar{V}^R_{t+1}(w_{t+1}, X_{t+1})$ for every $\alpha
>0$.

For $X_{t+1} > w_t \DA_{t+1}$, 
we then obtain by
\eqref{eqn:Ratchet_BcwdsRecursion_VRt}
that 
$V_{t+1}^R(w_t,X_{t+1}) =
\bar{V}_{t+1}^R(X_{t+1}/\DA_{t+1},X_{t+1}) =X_{t+1}\cdot
\bar{V}_{t+1}^R(1/\DA_{t+1},1)$. 
Hence $\partial_{X_{t+1}} V_{t+1}^R
\equiv \bar{V}_{t+1}^R(1/\DA_{t+1},1) =
\bar{V}_{t+1}^R(1,\DA_{t+1})/\DA_{t+1}$ and
\mbox{$\partial_{X_{t+1}}^2 V_{t+1}^R \equiv 0$.} In particular,
assumption \eqref{eqn:Ratchet_TechGrowthCondition} is satisfied
together with lemma \ref{lemma:Hopital}.
For any real number $x \leq \DA_{t+1}$, we deduce together with 
lemma \ref{lemma:ScalingProp_Derivaties} that $\partial_{X_{t+1}}^l
V_{t+1}^R(w_t,w_t x) = 
\partial_{X_{t+1}}^l \bar{V}_{t+1}^R(w_t,w_t x)=$ $w_t^{1-l}
\partial_{X_{t+1}}^l \bar{V}_{t+1}^R(1,x)$ for $l= 1,2$.

Summarizing all effects of the assumption that $\bar{V}^R_{t+1}$
satisfies (A2) allows to rewrite
\begin{eqnarray}
\label{eqn:Ratchet_Proof_BkwdRec_hatVRt}
\bar{V}^R_t(w_{t},X_t) &=&w_t \cdot \bar{V}_{t+1}^R(1,\DA_{t+1})\\
 &&+ \, C_t\left(w_t(1+\DA_{t+1}) | X_t\right) \cdot \bar{V}_{t+1}^R(1,\DA_{t+1})/\DA_{t+1}\nonumber\\
 &&- \, P_t\left(w_t(1+\DA_{t+1}) | X_t\right) \cdot \partial_{X_{t+1}}
   \bar{V}_{t+1}^R(1,\DA_{t+1})\nonumber\\
 &&+ \int_{1}^{1+\DA_{t+1}} P_t(w_t k|X_t)\cdot w_t^{-1} \cdot
\partial_{X_{t+1}}^2 \bar{V}_{t+1}^R(1,k-1) \cdot w_t \, dk \, .\nonumber
\end{eqnarray}
This representation shows that also $\bar{V}^R_t$ satisfies (A2),
i.e. $\bar{V}^R_{t}(\alpha \cdot w_{t}, \alpha \cdot X_t) = \alpha
\cdot \bar{V}^R_{t}(w_{t}, X_t)$ for every $\alpha >0$. Since the
terminal term
$\bar{V}^R_{N-1}=\bar{V}^R_{N-1}(w_{N-1},X_{N-1})=P_{N-1}(w_{N-1}|X_{N-1})$
clearly satisfies (A2) (as $S$ satisfies (A2) by assumption),
backwards induction shows that $\bar{V}^R_t$ satisfies (A2) for
every $t=0,\dots, N-1$.

Relation \eqref{eqn:Ratchet_Proof_BkwdRec_hatVRt} delivers the
identification \eqref{eqn:Ratchet_Identif_AlphBetaGRt} of the
constants $\alpha_t$,$\beta_t$ and  the weight functions $g^R_t$.
Differentiation of the backward recursions
\eqref{eqn:Ratchet_Proof_BkwdRec_hatVRt} applied to
$\bar{V}^R_{t+1}$ with respect to $X_{t+1}$ leads to the backwards
recursions \eqref{eqn:Ratchet_Def_AlphBetWeights_gt} for $\alpha_t$,
$\beta_t$ and $g^R_t$. \qed



\newpage

\begin{table}[!hp]
	\centering
		\begin{tabular}{l| r r| r r r}
&	\multicolumn{2}{l|}{{NPV}}	&	\multicolumn{3}{l}{{Sensitivity}}\\
\cline{2-6} 
&base	&HW vol 	&absolut&	in \% npv &	in \% sensi \\
&			& +0.1\%	&				&	base  		&		vanilla put\\
\hline
\multicolumn{6}{l}{10y}\\	
\hline				
vanilla put 	&	 8.108 &	 8.153 &	 0.045 &	0.55\%&\\	
withdrawal guarantee	& 5.122 	& 5.135 &	 0.013 &	0.26\% &	30.07\%\\
put on mlt-contrb. fund &	 4.754 &	 4.783 &	 0.028 &	0.60\% &	63.03\%\\
\hline
\multicolumn{6}{l}{20y}\\	 					
\hline
vanilla put	& 18.288 &	 18.603 &	 0.314 &	1.72\%	&\\
withdrawal guarantee &	 11.989 &	 12.090 &	 0.101 &	0.84\%	& 32.20\%\\
put on mlt-contrb. fund	& 10.903 &	 11.105 &	 0.203 &	1.86\%	& 64.46\%\\
\hline
\multicolumn{6}{l}{30y}\\	 										
\hline
vanilla put	& 27.809 &	 28.605 &	 0.796 &	2.86\% &\\	
withdrawal guarantee	& 19.088 &	 19.377 &	 0.289 &	1.52\%	& 36.32\%\\
put on mlt-contrb. fund	& 16.900 &	 17.437 &	 0.537 &	3.18\%	& 67.45\%\\
\end{tabular}
\caption{
Interest rate volatility sensitivity comparison for the following products with maturities of
10, 20 and 30 years: (a) vanilla put, (b) withdrawal guarantee and (c) put on multi-contribution fund.
The values are calculated under the hybrid one-factor Hull-White \& Black-Scholes model (2\% flat 
initial forward rate, 3\% mean reversion and 1\% volatility of short rate, 25\% excess voatility
of underlying with zero correlation to short rate). 
The regular withdrawals or contributions are assumed to occur quarterly with unit amount.
The guarantees are assumed at-the-money such that the initial fund value or the 
strikes are set equal to the sum of the regular withdrawals or contributions, respectively.
The vanilla put is valuated analytically, the other guarantees are priced 
with a Monte-Carlo approach (using $10^5$ scenarios).}
\label{tab:HWSensie}
\end{table}

\newpage

\newpage

\newpage

\newpage

\newpage

\newpage

\vspace{-2cm}

\newpage

\newpage

\end{appendix}

\end{document}